\documentclass[12pt,a4paper]{article}
\usepackage{amsmath}
\usepackage[font={footnotesize,it}]{caption}
\usepackage[top=1.1in, bottom=1.1in, left=1.1in, right=1.1in]{geometry}

\DeclareCaptionStyle{italic}[justification=centering]{labelfont={bf},textfont={it},labelsep=colon}
\usepackage{graphicx,psfrag,epsf}
\usepackage{enumerate}
\usepackage{natbib}
\usepackage{url, setspace}
\usepackage{booktabs, subfig, bm, paralist,mathpazo,tikz,todonotes,longtable,microtype,dsfont,rotating} 
\usepackage[pdftex,colorlinks=true]{hyperref}
\definecolor{darkblue}{rgb}{0,0,.6}
\hypersetup{citecolor=darkblue,linkcolor=darkblue,urlcolor=darkblue}

\newcommand{\blind}{0}

\graphicspath{{plots/}}
\DeclareMathOperator*{\argmin}{\arg\!\min}
\newsavebox\CBox
\def\textBF#1{\sbox\CBox{#1}\resizebox{\wd\CBox}{\ht\CBox}{\textbf{#1}}}

\date{\today}

\begin{document}

\def\spacingset#1{\renewcommand{\baselinestretch}
{#1}\small\normalsize} \spacingset{1}

\if0\blind
{
  \title{\bf Dynamic principal component regression: Application to age-specific mortality forecasting}
  \author{Han Lin Shang\thanks{Postal address: Research School of Finance, Actuarial Studies and Statistics, Level 4, Building 26C, Australian National University, Kingsley Street, Acton, Canberra, ACT 2601, Australia; Telephone: +61(2) 612 50535; Fax: +61(2) 612 50087; Email: hanlin.shang@anu.edu.au.}
  \hspace{.2cm}\\
    Research School of Finance, Actuarial Studies and Statistics \\
    Australian National University}
  \maketitle
} \fi

\if1\blind
{
   \title{\bf Dynamic principal component regression: Application to age-specific mortality forecasting}
   \author{}
   \maketitle
} \fi

\bigskip

\begin{abstract}
In areas of application, including actuarial science and demography, it is increasingly common to consider a time series of curves; an example of this is age-specific mortality rates observed over a period of years. Given that age can be treated as a discrete or continuous variable, a dimension reduction technique, such as principal component analysis, is often implemented. However, in the presence of moderate to strong temporal dependence, \textit{static} principal component analysis commonly used for analyzing independent and identically distributed data may not be adequate. As an alternative, we consider a \textit{dynamic} principal component approach to model temporal dependence in a time series of curves. Inspired by \citeauthor{Brillinger74}'s \citeyearpar{Brillinger74} theory of dynamic principal components, we introduce a dynamic principal component analysis, which is based on eigen-decomposition of estimated long-run covariance. Through a series of empirical applications, we demonstrate the potential improvement of one-year-ahead point and interval forecast accuracies that the dynamic principal component regression entails when compared with the static counterpart.
\\

\end{abstract}

\newpage
\spacingset{1.44}

\section{Introduction}

In many developed countries, increases in longevity and an aging population have led to concerns regarding the sustainability of pensions, healthcare, and aged-care systems \citep[e.g.,][]{OECD13}. These concerns have resulted in a surge of interest among government policymakers and planners to engage in accurate modeling and forecasting of age-specific mortality rates. In addition, forecasted mortality rates are an important input for determining annuity prices and thus are very important to pension and insurance industries \citep[see, e.g.,][]{SH17b}. Many statistical methods have been proposed for forecasting age-specific mortality rates \citep[for reviews, see][]{BT08}. Of these, a significant milestone in demographic forecasting was the work by \cite{LC92}. They implemented a principal component method to model age-specific mortality rates and extracted a single time-varying index of the level of mortality rates, from which the forecasts were obtained by a random walk with drift. 

The strengths of the Lee--Carter (LC) method are its simplicity and robustness in situations where age-specific log mortality rates have linear trends \citep{BHT+06}. The main weakness of the LC method is that it attempts to capture the patterns of mortality rates using only one principal component and its associated scores. To rectify this deficiency, the LC method has been extended and modified. For example, from a time series of matrix perspective, \cite{RH03} proposed the use of more than one component in the LC method to model age-specific mortality. From a time series of function perspective, \cite{HU07} proposed a functional time-series method that uses nonparametric smoothing and higher-order principal components.

A common feature of the aforementioned works is that a static principal component analysis (PCA) is often used to decompose a time series of data matrix or curves. Under moderate to strong temporal dependence, the extracted principal components may not be consistent because of temporal dependence, leading to erroneous estimators. To overcome this issue, we consider a dynamic approach that extracts principal components based on an estimated long-run covariance instead of estimated variance alone. Note that the long-run covariance includes the variance function as a component, yet also measures temporal cross-covariance at different positive and negative lags. Similar to the finite-dimensional time-series framework, the long-run covariance estimation is the sum of empirical autocovariance functions and is often truncated at some finite lag in practice (see Section~\ref{sec:2}).

While the LC method is commonly used for analyzing mortality rates at discrete ages, the functional time-series method is often used for analyzing mortality curves where age is treated as a continuum. With these two methods, the contribution of this paper is to demonstrate the improvement of point and interval forecast accuracies that the dynamic principal component regression entails when compared with the static PCA for modeling and forecasting age-specific mortality rates at a one-year-ahead forecast horizon. In the longer forecast horizon, the difference in forecast accuracy becomes marginal, and these results can be obtained upon request from the author.
 
The rest of this paper is structured as follows. In Section~\ref{sec:2}, we describe a kernel sandwich estimator for estimating long-run covariance. Based on the estimated long-run covariance, we introduce an eigen-decomposition that extracts dynamic principal components and their associated scores in Section~\ref{sec:3}. Illustrated by empirical data obtained from the \cite{HMD17} in Section~\ref{sec:4}, we evaluate and compare the one-year-ahead point and interval forecast accuracies between the LC and functional time-series methods described in Section~\ref{sec:forecasting_methods}, using both the static and dynamic principal component regression models in Section~\ref{sec:results}. Conclusions are given in Section~\ref{sec:5}.

\section{Datasets}\label{sec:4}

The datasets used in this study were taken from the \cite{HMD17}. For each sex in a given calendar year, the mortality rates obtained by the ratio between ``number of deaths" and ``exposure to risk" were arranged in a matrix for age and calendar year. Twenty-four countries, mainly developed nations, were selected, and thus 48 sub-populations of age- and sex-specific mortality rates were obtained for all analyses. The 24 countries selected all had reliable data series commencing during or before 1950. As a result of possible structural breaks (i.e., two world wars), we truncated all data series from 1950 onwards. The omission of Germany was because the Human Mortality Database for a reunited Germany only goes back to 1990. The selected countries are shown in Table~\ref{tab:1}, alongside their final year of available data. To avoid fluctuation in older ages, we considered ages from 0 to 99 in a single year of age, and the last age group was from 100 onwards.

\begin{table}[!htbp]
\def\arraystretch{1.2}\tabcolsep 0.12in
\centering
\caption{\small The 24 countries examined in this study, with the initial year of 1950 and their final year listed below.}\label{tab:1}
\begin{small}
\begin{tabular}{@{}llllll@{}}
\toprule
Country &  Abbreviation & Final year & Country & Abbreviation & Final year\\
\midrule
Australia & AUS & 2014  & Italy & ITA & 2012 \\
Austria & AUT & 2014  & Japan & JPN & 2014 \\
Belgium & BEL & 2015 & The Netherlands & NLD & 2014 \\
Bulgaria & BGR & 2010 & Norway & NOR & 2014  \\
Canada & CAN & 2011  & New Zealand & NZ & 2013  \\
The Czech Republic & CZE & 2014 & Portugal & PRT & 2015  \\
Denmark & DEN & 2014 & Spain & SPA & 2014 \\
Finland & FIN & 2015 & Slovakia & SVK & 2014 \\
France & FRA & 2014 & Sweden & SWE & 2014 \\
Hungary & HUN & 2014 & Switzerland & SWI & 2014  \\
Iceland & ICE & 2013 & The United Kingdom & UK & 2013  \\
Ireland & IRL & 2014 & The United States & US & 2015 \\
\bottomrule
\end{tabular}
\end{small}
\end{table}

\subsection{Functional time-series plot}\label{sec:4.1}

\begin{figure}[!htbp]
\centering
\subfloat[Original series]
{\includegraphics[width=8.7cm]{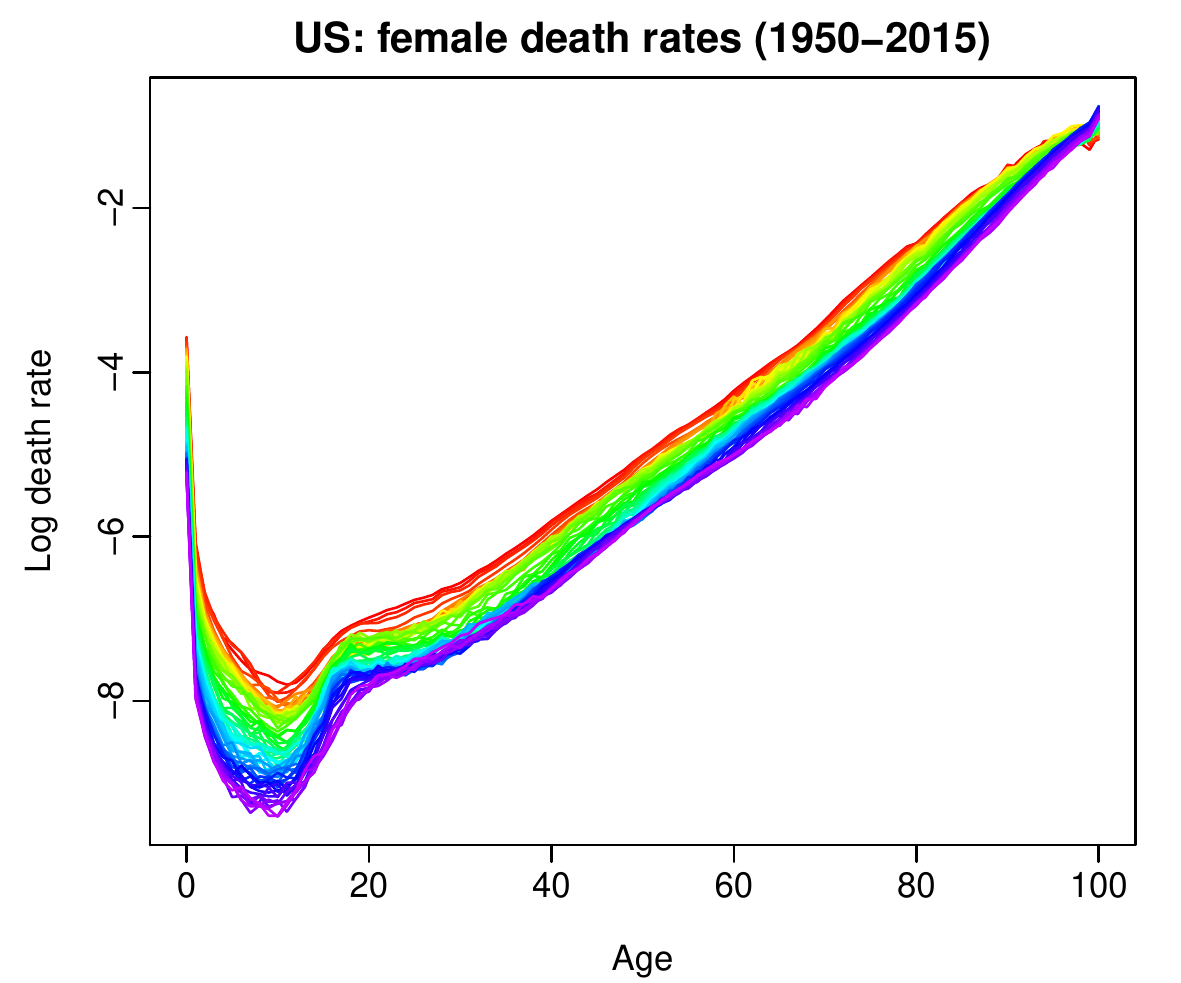}\label{fig:1a}}
\quad
\subfloat[Smoothed series]
{\includegraphics[width=8.7cm]{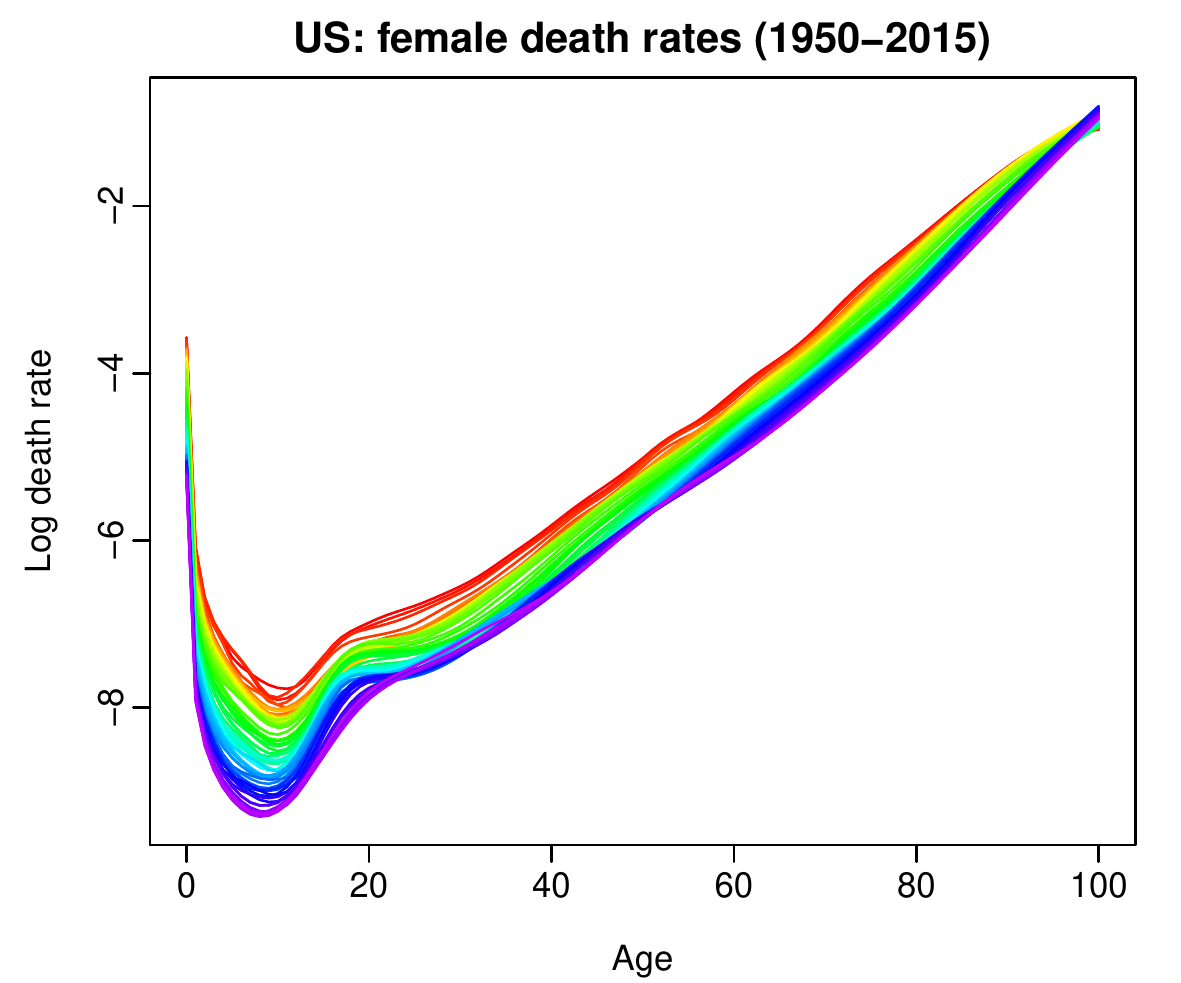}\label{fig:1b}}
\caption{\small Observed and smoothed age-specific female log mortality rates in the US. Data from the distant past are shown in red, and the most recent data are shown in purple.}\label{fig:1}
\end{figure}

To present an evolution of age-specific mortality, we present a functional time-series plot for the raw female log mortality rates in the US in Figure~\ref{fig:1a}, while the functional time-series plot for the smoothed data is shown in Figure~\ref{fig:1b}. 

To smooth these functional time series, we assumed there was an underlying $L_2$ continuous and smooth function $f_t(x)$, such that
\begin{equation*}
\mathcal{Y}_t(x_j) = f_t(x_j)+\sigma_t(x_j)\epsilon_{t,j}, \qquad j=1,\dots,p, \quad t=1,\dots,n,
\end{equation*}
where $\mathcal{Y}_t(x_j)$ denotes the raw log mortality rates, $f_t(x_j)$ denotes the smoothed log mortality rates, $\{\epsilon_{t,j}\}$ represents independent and identically distributed (IID) random variables across $t$ and $j$ with a mean of zero and a unit variance, and $\sigma_t(x_j)$ allows for heteroskedasticity and can be estimated by
\begin{equation*}
\widehat{\sigma}_t(x_j) = \frac{1}{\exp\{\mathcal{Y}_t(x_j)\}\text{E}_t(x_j)},
\end{equation*}
where $\text{E}_t(x_j)$ denotes population of age $x_j$ at June 30 in year $t$ (often known as the ``exposure-at-risk").

Given that the log mortality rates increased linearly over age, we used a penalized regression spline with monotonic constraint, where the monotonicity was imposed for ages at and above 65 \citep[for details, see][]{HU07}. With the weights equal to the inverse variances $w_t(x_j) = 1/\widehat{\sigma}_t^2(x_j)$, the smoothed log mortality rate was obtained by
\begin{equation*}
f_t(x_j) = \argmin_{\theta_t(x_j)}\sum^M_{j=1}w_t(x_j)|\mathcal{Y}_t(x_j) - \theta_t(x_j)|+\lambda\sum^{M-1}_{j=1}|\theta_t^{'}(x_{j+1}) - \theta_t^{'}(x_j)|,
\end{equation*}
where $x_j$ represents different ages (grid points) in a total of $M$ grid points, $\lambda$ denotes a smoothing parameter, $\theta^{'}$ denotes the first derivative of smooth function $\theta$, which can both be approximated by a set of B-splines.

Figure~\ref{fig:1} is an example of the rainbow plot, where the colors of the curves follow the order of a rainbow with the oldest data shown in red and most recent data shown in violet \citep[see also][]{HS10}. By analyzing the changes in mortality as a function of both age $x$ and year $t$, it can be seen that mortality rates showed a gradual decline over the years. Mortality rates dipped from their early childhood high, climbed in the teen years, stabilized in the early twenties, and then steadily increased with age. We further noted that, for both males and females, log mortality rates declined over time, especially in the younger and older ages. 


\subsection{Mortality improvement rate}

In demography and actuarial science, a time series of age-specific mortality rates is commonly modeled and forecast at a logarithmic scale. These series are non-stationary, as the mean function changes over time. As an alternative approach, we can model the improvement in mortality rates, rather than the rate itself \citep[see, e.g.,][]{HR12}. The advantage of modeling the mortality improvement is that the data series is stationary. As one way of measuring mortality improvement, the year-on-year mortality improvement rate of \cite{HR12} was considered and expressed as
\begin{equation}
z_{x,t} = 2 \times \frac{1- m_{x,t}/m_{x,t-1}}{1 + m_{x,t}/m_{x,t-1}} = 2\times \frac{m_{x,t-1}-m_{x,t}}{m_{x,t-1}+m_{x,t}},\qquad t=2,\dots,n, \label{eq:improvement}
\end{equation}
for age $x$ in year $t$, where $m_{x,t}$ denotes the raw mortality rate, $z_{x,t}$ denotes the transformed mortality rate and $n$ symbolizes the number of years. 

The expression in Eq.~\eqref{eq:improvement} can be seen as the ratio between the incremental mortality improvement $(m_{x,t-1} - m_{x,t})$ and the average $(m_{x,t} + m_{x,t-1})/2$ of two adjacent mortality rates. By defining the denominator of the ratio in this way, we avoided the small phase difference between the numerator and denominator that would otherwise be the case. Thus, improving incremental mortality rate changes implied $z_{x,t}>0$, while deteriorating incremental changes implied $z_{x,t}<0$.

Via back-transformation of Eq.~\eqref{eq:improvement}, we obtained:
\begin{equation*}
m_{x,t} = \frac{2+z_{x,t}}{2-z_{x,t}}\times m_{x,t-1}.
\end{equation*}

\begin{figure}[!htbp]
\centering
\subfloat[Original series]
{\includegraphics[width=8.7cm]{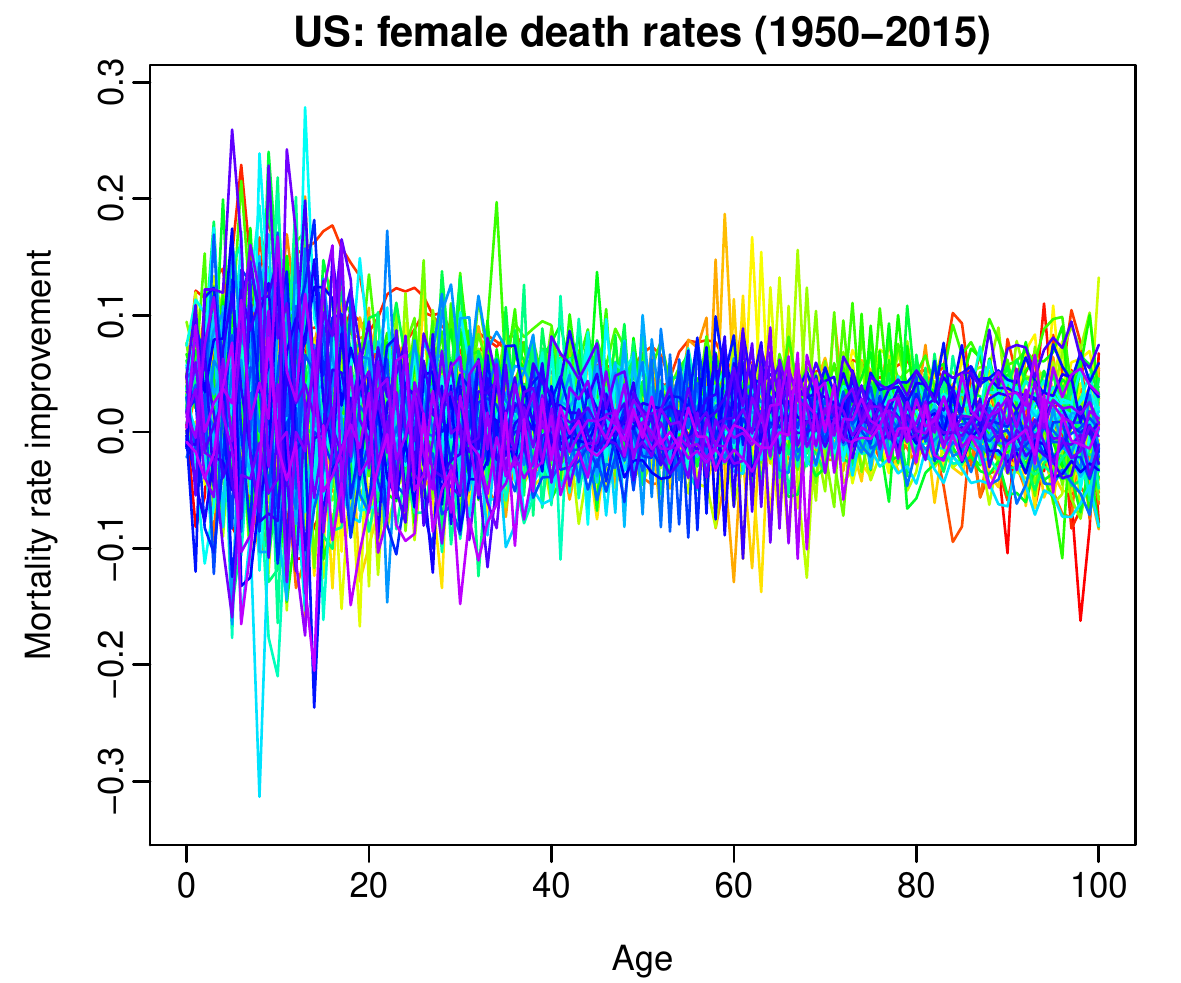}\label{fig:20a}}
\quad
\subfloat[Smoothed series]
{\includegraphics[width=8.7cm]{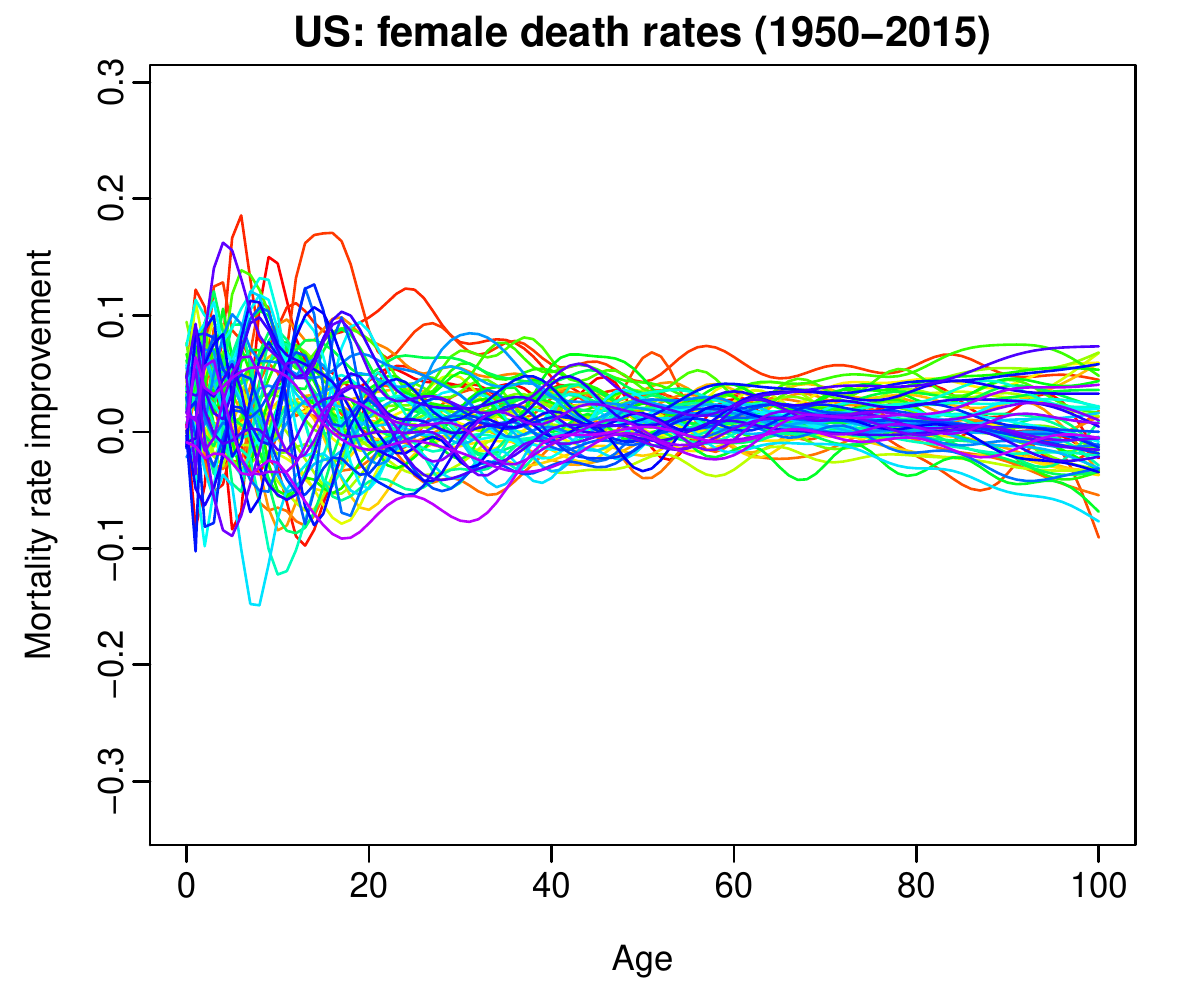}\label{fig:20b}}
\caption{\small Observed and smoothed age-specific female mortality rate improvements in the US.}
\end{figure}

In Figures~\ref{fig:20a} and~\ref{fig:20b}, we plot the observed and smoothed curves for the age-specific female mortality rate improvements in the US. The curves are stationary and more volatile in the early ages (i.e., ages between 0 and 40) than the later ages. We obtained smoothed mortality rate improvement by computing the smoothed age-specific mortality curve, and then applying Eq.~\eqref{eq:improvement}.

\section{Forecasting methods}\label{sec:forecasting_methods}

Given that the focus of this paper is a comparison of short-term forecast accuracy between the static and dynamic principal component analyses, we revisited the LC and functional time-series methods as two possible methods for forecasting age-specific mortality rates. The LC model considers age a discrete variable, while the functional time-series model treats age as a continuous variable. We denoted with $m_{x,t}$ the observed mortality rate at age $x$ in year $t$, calculated as the number of deaths aged $x$ in year $t$, divided by the corresponding mid-year population aged $x$ in year $t$. With $m_{x,t}$, we first obtained transformed series $z_{x,t}$ from Eq.~\eqref{eq:improvement}. 

\subsection{Adapted LC method}

The original LC method was applied to model the log mortality rate \citep{LC92}. Here, we extended it to model mortality improvement rate. The formulation of our adapted LC model is given by
\begin{equation}
z_{x,t} = a_x + b_x \kappa_t + \epsilon_{x,t}, \label{eq:1}
\end{equation}
where $a_x$ denotes the age pattern of the mortality rates averaged over years, $b_x$ denotes the first principal component at age $x$, $\kappa_t$ denotes the first set of principal component scores at year $t$ and measures the general level of the mortality rates, and $\epsilon_{x,t}$ denotes the residual at age $x$ and year $t$.

The LC model in Eq.~\eqref{eq:1} is over-parametrized, in that the model structure is invariant under the following transformations:
\begin{align*}
\{a_x, b_x, \kappa_t\}  &\mapsto \{a_x, b_x/c, c\kappa_t\}, \\
\{a_x, b_x, \kappa_t\}  &\mapsto \{a_x - c b_x, b_x, \kappa_t + c\}
\end{align*}
To ensure the model identifiability, \cite{LC92} imposed two constraints given as
\begin{equation*}
\sum^n_{t=1}\kappa_t = 0, \qquad \sum_{x=x_1}^{x_p}b_x = 1,
\end{equation*}
where $n$ denotes the number of years and $p$ denotes the number of ages in the observed dataset.

Instead of using a random walk with drift, the set of principal component scores, $\kappa_t$, can be extrapolated using autoregressive integrated moving average (ARIMA) models. We used the automatic algorithm of \cite{HK08} to choose the optimal orders of autoregressive $p$, moving average $q$, and difference order $d$. $d$ was selected based on successive Kwiatkowski--Phillips--Schmidt--Shin (KPSS) unit root tests \citep{KPSS92}. KPSS tests were used to test the null hypothesis that an observable time series was stationary around a deterministic trend. We tested the original data (i.e., the first set of principal component scores) for a unit root; if the test result was significant, then we tested the differenced data for a unit root. The procedure continued until we obtained our first insignificant result. Having determined $d$, the orders of $p$ and $q$ were selected based on the optimal Akaike information criterion with a correction for finite sample sizes \citep{Akaike74}. After identifying the optimal ARIMA model, the maximum likelihood method could then be used to estimate the parameters. Conditioning on the estimated mean, the estimated first principal component $\widehat{b}_x$, and the observed mortality rate improvement, the $h$-step-ahead point forecast of $z_{x,n+h}$ can be expressed as:
\begin{equation*}
\widehat{z}_{x,n+h|n} = \text{E}\left[z_{x,n+h}\Big|z_{x,1},\dots,z_{x,n}, \widehat{a}_x, \widehat{b}_x\right] = \widehat{a}_x + \widehat{b}_x \widehat{\kappa}_{n+h|n},
\end{equation*}
where $\widehat{a}_x = \frac{1}{n}\sum_{t=1}^nz_{x,t}$ denotes the estimated mean, and $\widehat{\kappa}_{n+h|n}$ denotes the $h$-step-ahead forecast of the principal component scores.

In \cite{HR12}, the LC method does not include the mean $a_x$, since their generalized linear model approach uses a Newton-Raphson iterative fitting algorithm to estimate $b_x$ and $\kappa_t$ by minimizing a deviance criterion. In contrast, we applied the PCA to mortality improvement $z_{x,t}$. The PCA often requires de-centering the data. In Figure~\ref{fig:point}, we also compare the one-step-ahead forecast performances of the LC method under a Poisson error structure without centering. We found that the difference regarding whether or not to center the data was marginal in terms of forecast accuracy. In Appendix B, we also compare the five-step-ahead and 10-step-ahead forecast accuracy using the LC method with and without centering.

\subsection{Functional time-series method}

Functional time series often consist of random functions observed at regular time intervals. In the context of mortality, functional time series can arise when observations in a time period can be considered together as finite realizations of an underlying continuous function \citep[e.g.,][]{HU07}. There are several advantages to consider the functional time-series method:
\begin{enumerate}
\item[1)] Data points may be observed sparsely. Via the functional time-series method, the underlying trajectory may be recovered \citep[for details, see][]{ZW16}.
\item[2)] With continuity, derivative information can provide new insight into data analysis \citep[see, e.g.,][]{Shang17}. 
\item[3)] A nonparametric smoothing technique can be incorporated into the modeling procedure to obtain smoothed principal components. Smoothing deals with one criticism of the LC model; namely, that the estimated values, $b_x$, can be subject to considerable noise and, without smoothing, this would be propagated into forecasts of future mortality rates. Smoothing can reduce measurement error and increase the signal-to-noise ratio, and also deals with estimating missing values for some ages at a given year.
\item[4)] Given that the functional time-series method can consider more than one component, \cite{Shang12} indicated that the functional time-series method outperforms the LC method.
\end{enumerate}

Many possible nonparametric smoothing techniques have been proposed, such as basis spline \citep[for details, see][]{deBoor01}. We used a penalized regression spline with a monotonic constraint \citep[for details, see][]{HU07}. The smoothed log mortality rates $\bm{f}(x) = \{f_1(x),\dots, f_n(x)\}$ were treated as realizations of a stochastic process. In \cite{HU07}, they considered modeling the smoothed log mortality rates directly. Given that the log mortality rates are non-stationary, we considered modeling and forecasting mortality rate improvement. From Eq.~\eqref{eq:improvement}, we could obtain a set of transformed and smoothed series, denoted by $\bm{z}(x) = \{z_1(x),\dots,z_n(x)\}$. Using functional PCA, these smoothed mortality improvement curves were decomposed into
\begin{equation}
z_t(x) = a(x) + \sum_{k=1}^Kb_k(x)\kappa_{t,k} + e_t(x), \qquad t=1,\dots,n, \label{eq:2}
\end{equation}
where $a(x)$ denotes the mean function estimated by $\widehat{a}(x) = \frac{1}{n}\sum^n_{t=1}z_t(x)$, $\left\{b_1(x),\dots,b_K(x)\right\}$ denotes a set of functional principal components, $\{\kappa_{t,1},\dots,\kappa_{t,K}\}$ denotes a set of principal component scores in year $t$, $e_t(x)$ denotes the error function with mean zero, and $K<n$ denotes the number of principal components retained. 

Decomposition in Eq.~\eqref{eq:2} facilitates dimension reduction because the first $K$ terms often provide a reasonable approximation to the infinite sums; thus, the information contained in $\bm{z}(x)$ can be adequately summarized by the $K$-dimensional vector, $\bm{\Phi} = \left[\bm{b}_1(x),\dots,\bm{b}_K(x)\right]$. In contrast to the LC model, another advantage of the functional time-series model is that more than one component may be used to improve model fitting \citep[see also][]{RH03}. Here, the number of components is determined as the minimum that reaches a certain level of the proportion of total variance explained by the leading components, such that
\begin{equation}
K=\argmin_{K: K\geq 1}\left\{\sum^K_{k=1}\widehat{\lambda}_k\Big/\sum^{\infty}_{k=1}\widehat{\lambda}_k\mathds{1}_{\left\{\widehat{\lambda}_k>0\right\}}\geq 85\%\right\}, \label{eq:ncomp}
\end{equation}
where $\widehat{\lambda}_k$ represents the $k$\textsuperscript{th} estimated eigenvalue, and $\mathds{1}_{\left\{\widehat{\lambda}_k>0\right\}}$ is to exclude possible zero eigenvalues, and $\mathds{1}\{\cdot\}$ represents the binary indicator function. The threshold of 85\% is advocated in \citet[][p.41]{HK12}.

Conditioning on the estimated mean function $\widehat{a}(x)$, the estimated functional principal components $\bm{\Phi}$, and the observed mortality rate improvement $\bm{z}(x)$, the $h$-step-ahead point forecast of $z_{n+h}(x)$ can be expressed as
\begin{align*}
\widehat{z}_{n+h|n}(x) &= \text{E}\left[z_{n+h}(x)|\bm{z}(x),\widehat{a}(x),\bm{\Phi}\right] \\
&= \widehat{a}(x) + \sum^K_{k=1}\widehat{b}_k(x)\widehat{\kappa}_{n+h|n,k},
\end{align*}
where $\widehat{a}(x)$ denotes the estimated mean function, $\widehat{b}_k(x)$ denotes the $k$\textsuperscript{th} estimated functional principal component, and $\widehat{\kappa}_{n+h|n,k}$ denotes the $k$\textsuperscript{th} estimated principal component scores obtained via a univariate or multivariate time-series forecasting method. Given that it can handle non-stationarity, we considered a univariate forecasting method, such as the ARIMA model with orders selected automatically. 

The critical component of the aforementioned forecasting methods is the static PCA which was designed for IID data. In the presence of moderate to strong dependent data, the static PCA is not optimal because it does not incorporate autocovariance at different lags in a functional time series. As an alternative, we introduced a dynamic principal component analysis (DPCA) constructed from an eigen-decomposition of an estimated long-run covariance. The long-run covariance included the variance and autocovariance at lags greater than zero. 

\section{Long-run covariance and its estimation}\label{sec:2}

In statistics, long-run covariance enjoys vast literature in the case of finite-dimensional time series, beginning with the seminal work of \cite{Brillinger74}, and is still the most commonly used technique for smoothing the periodogram by employing a smoothing weight function and a bandwidth parameter. In the functional time series, long-run covariance plays an important role in modeling temporal dependence \citep[see e.g.,][]{RS16}.

To provide a formal definition of the long-run covariance, let $\{z_t(x)\}_{t\in Z}$ be a stationary and ergodic functional time series. For example, $z_t(x)$ could be used to denote the density of pollutants in a given city on day $t$ at intraday time $x$ or the mortality rate in year $t$ at age $x$. If $z_t(x)$ is non-stationary, it could be suitably transformed, so that the stationarity assumption holds. For a stationary functional time series, the long-run covariance is defined as:
\begin{align*}
C(x,u) &= \sum^{\infty}_{\ell = -\infty} \gamma_l(x,u) \\
&= \sum^{\infty}_{\ell=-\infty} \text{cov}\left[z_0(x), z_{\ell}(u)\right].
\end{align*}
Given that $\gamma_{\ell}(x,u)$ is symmetric and non-negative definite for any $\ell$, $C(x,u)$ is also symmetric and non-negative definite. By applying eigen-decomposition to the long-run covariance, $C(x, u)$, we obtained a set of eigenvalues and eigenfunctions.
 
In practice, we needed to estimate $C$ from a finite sample $\bm{z}(x) = [z_1(x),\dots, z_n(x)]$. Given its definition as a bi-infinite sum, a natural estimator of $C$ is:
\begin{equation}
\widehat{C}_{h,q}(x,u) = \sum^{\infty}_{\ell = -\infty}W_q\left(\frac{\ell}{h}\right)\widehat{\gamma}_{\ell}(x,u), \label{eq:11}
\end{equation}
where $h$ is called the bandwidth parameter, and
\begin{align*}
   \widehat{\gamma}_\ell(x,u)=\left\{
     \begin{array}{lr}
      \displaystyle \frac{1}{n}\sum_{j=1}^{n-\ell}\left[z_j(x)-\overline{z}(x)\right]\left[z_{j+\ell}(u)-\overline{z}(u)\right],\quad &\ell \ge 0
      \vspace{.3cm} \\
     \displaystyle \frac{1}{n}\sum_{j=1-\ell}^{n}\left[z_j(x)-\overline{z}(x)\right]\left[z_{j+\ell}(u)-\overline{z}(u)\right],\quad &\ell < 0,
     \end{array}
   \right.
\end{align*}
is an estimator of $\gamma_l(x,u)$, and $W_q$ is a symmetric weight function with bounded support of order $q$. The estimator in Eq.~\eqref{eq:11} was introduced in \cite{HK12} and \cite{RS16}, among others. As with the kernel estimator, the crucial part is the estimation of bandwidth parameter $h$. It can be selected through a data-driven approach, such as the plug-in algorithm proposed in \cite{RS16}. In Appendix A, we briefly describe the plug-in algorithm. With the estimated long-run covariance, we could obtain dynamic functional principal components and their scores, as described in Section~\ref{sec:3}.

\subsection{Application to US age-specific mortality rates} 

\begin{figure}[!htbp]
\centering
\subfloat[Sample long-run covariance for US female mortality]
{\includegraphics[width=8.5cm]{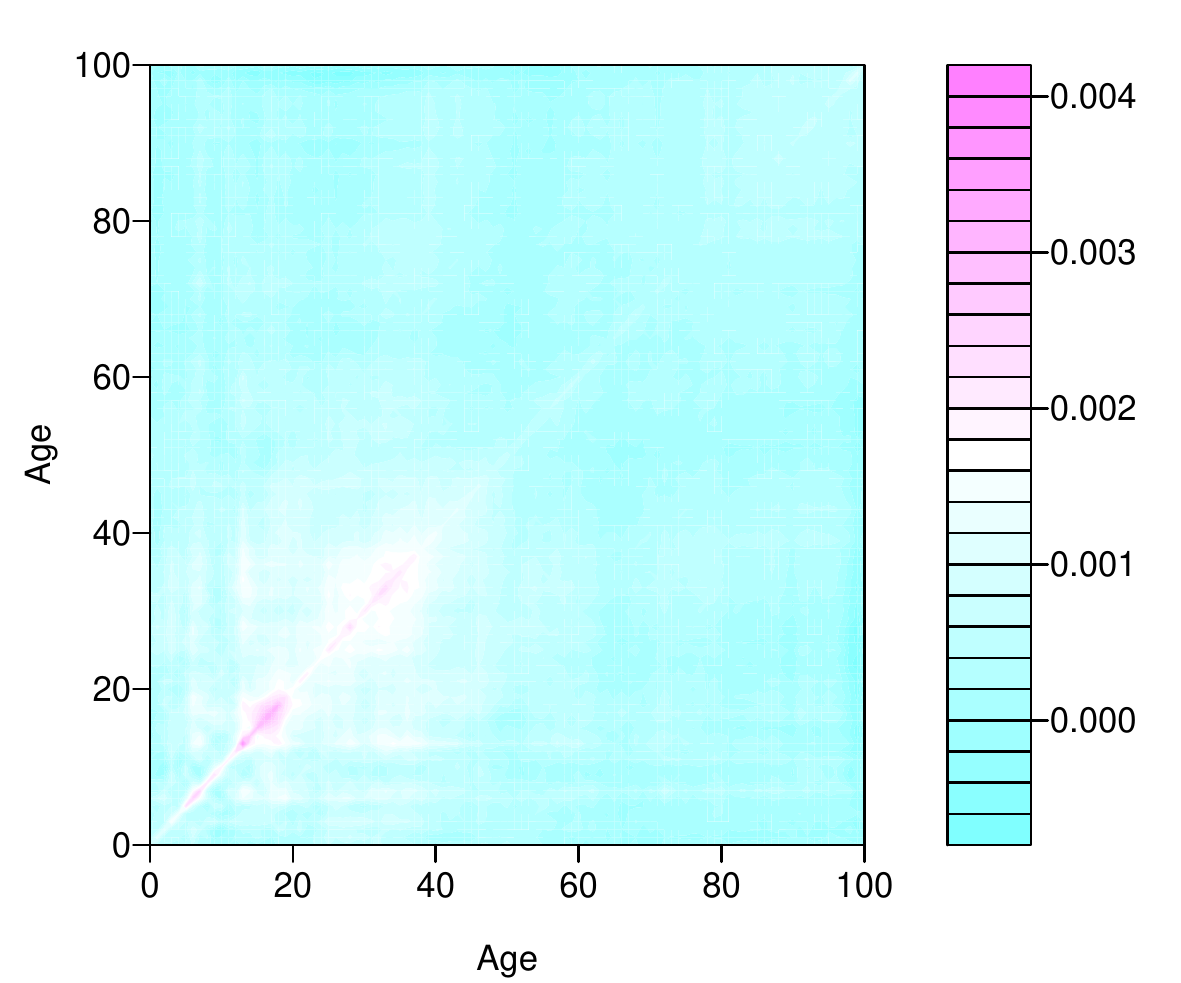}}
\subfloat[Sample variance for US female mortality]
{\includegraphics[width=8.5cm]{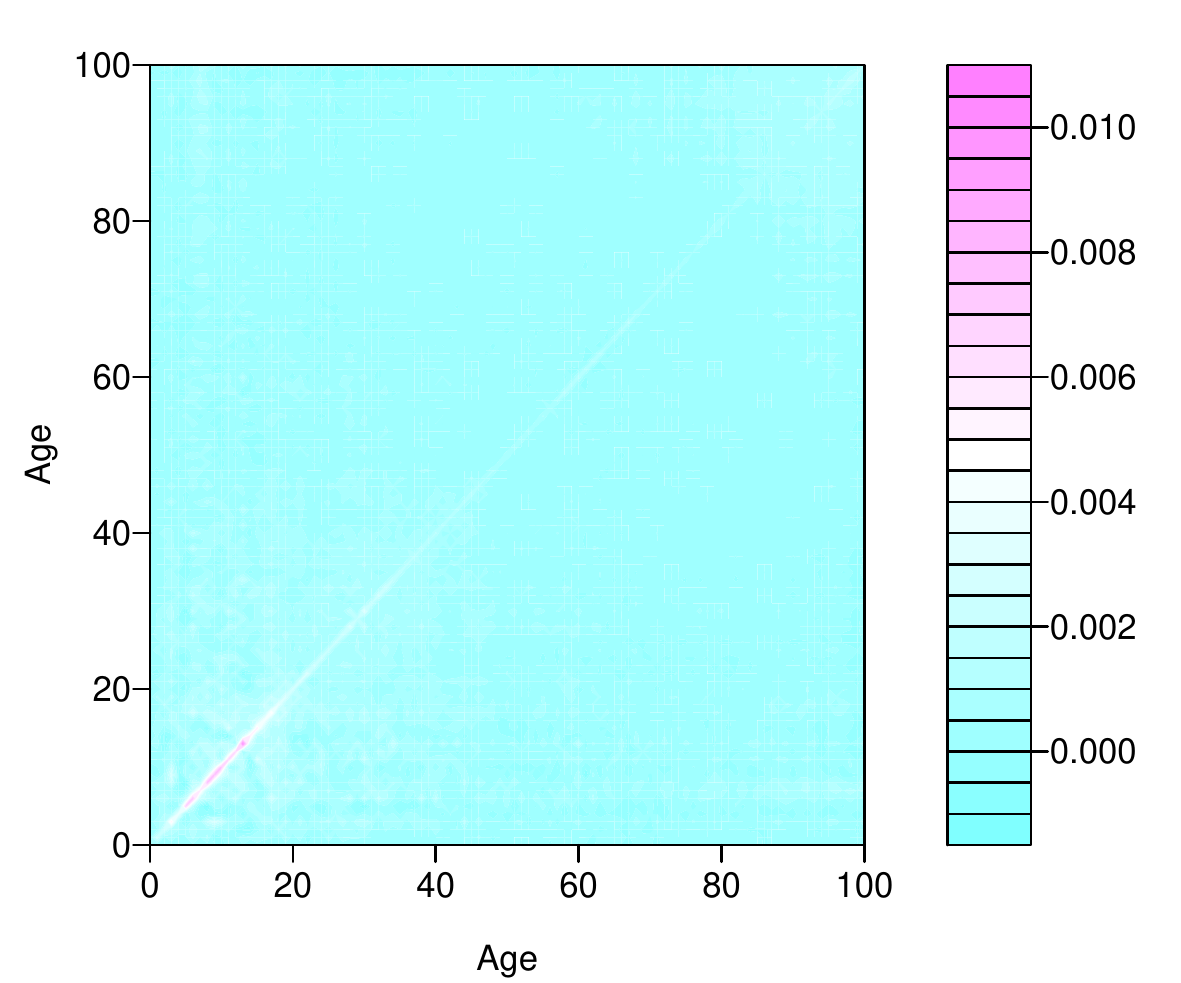}}
\caption{\small Comparison of sample long-run covariance and sample variance for US original female mortality rates.}\label{fig:2}
\end{figure}

Figure~\ref{fig:2} presents the estimated long-run covariance and variance for the female raw mortality rates in the US. With the input data as age-specific mortality improvement over years, we computed the sample long-run covariance and sample variance for a data matrix. The sample variance was computed by multiplying the data matrix by its transpose, while the sample long-run covariance was computed by the kernel sandwich estimator. Compared with the sample variance, the sample long-run covariance based on the kernel sandwich estimator with plugged-in bandwidth could incorporate an autocovariance structure, particularly for ages between 0 and 40. The mortality rate at the young ages exhibited higher variance than the mortality rate at the other ages. 

\begin{figure}[!htbp]
\centering
\subfloat[Sample long-run covariance for US smoothed female mortality]
{\includegraphics[width=8.5cm]{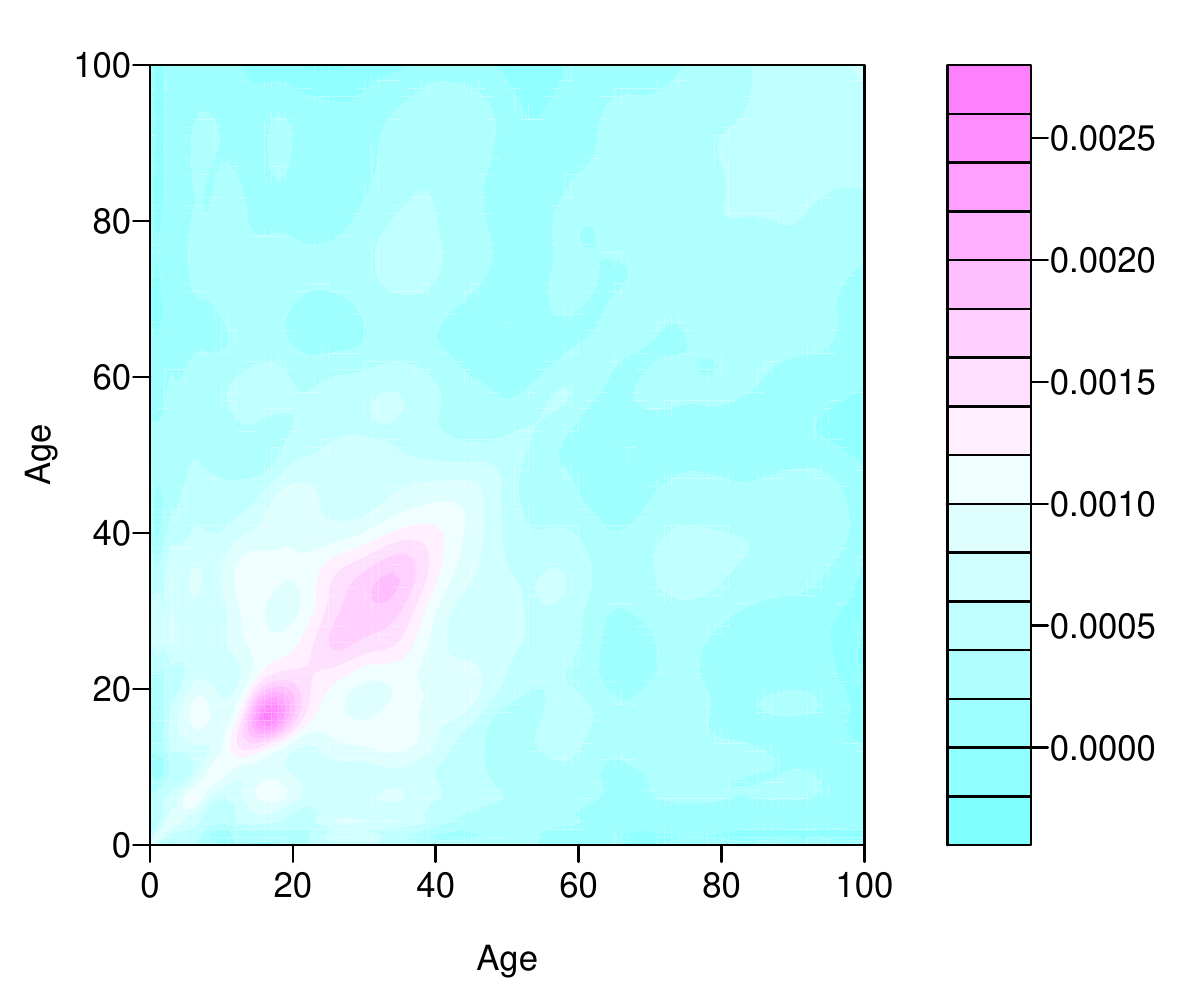}}
\subfloat[Sample variance for US smoothed female mortality]
{\includegraphics[width=8.5cm]{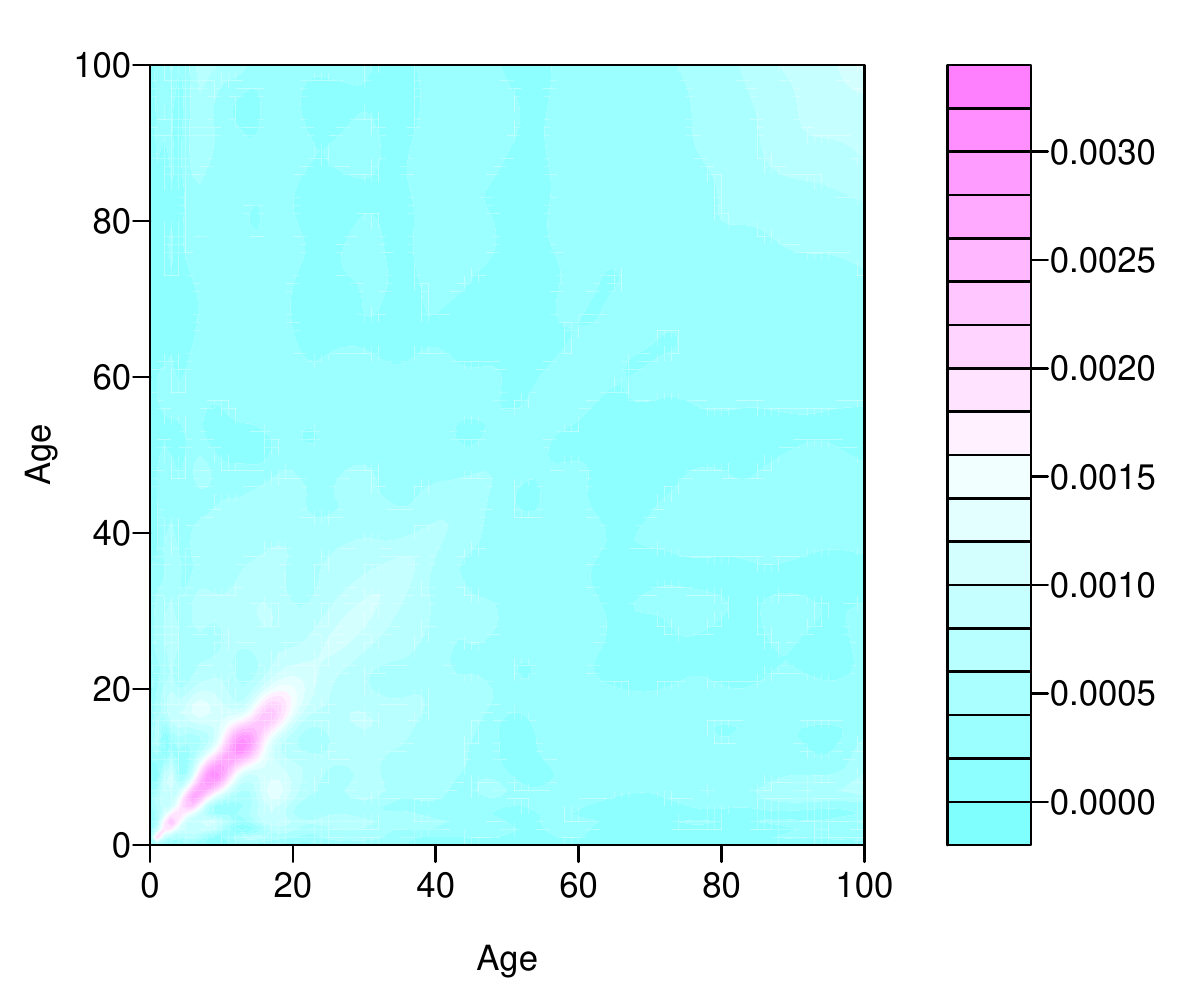}}
\caption{\small Comparison of sample long-run covariance and sample variance for US female mortality rates.}\label{fig:2_add}
\end{figure}

Compared with the long-run covariance based on the raw data series, the estimated long-run covariance based on the smoothed data series was smoother and showed a more explicit data structure in Figure~\ref{fig:2_add}. For estimating the long-run covariance, the estimated optimal bandwidth was 4.07 for the raw female data and 4.10 for the smooth female data.

The computation of long-run covariance relies heavily on the fact that our mortality improvement data were stationary time series. When the temporal dependence was weak, variance could be sufficient to estimate the long-run covariance. When the temporal dependence was moderate or high, one should include more terms in the long-run covariance estimation. The inclusion of autocovariance could improve forecast accuracy. From Figures~\ref{fig:2} and~\ref{fig:2_add}, it is clear that the long-run covariance also included the autocovariance at various lags.
  
\section{Dynamic functional principal component analysis}\label{sec:3}
 
From the long-run covariance, we applied functional principal component decomposition to extract the functional principal components and their associated scores. Via the Karhunen-Lo\`{e}ve expansion, a stochastic process, $z$, can be expressed as:
\begin{equation*}
z(x) = a(x) + \sum^{\infty}_{k=1}\beta_k\phi_k(x),
\end{equation*}
where $z^c(x) = z(x) - a(x)$. The principal component scores, $\beta_k$, are given by the projection of $z^c(x)$ in the direction of the $k$\textsuperscript{th} eigenfunction $\phi_k$ --- that is, $\beta_k = \langle z^c(x), \phi_k(x)\rangle$. The scores constitute an uncorrelated sequence of random variables with zero mean and variance $\lambda_k$. They can be interpreted as the weights of the contribution of the functional principal components $\phi_k(x)$ to $z^c(x)$.
 
Given that the long-run covariance, $C(x, u)$, is unknown, the population eigenvalues and eigenfunctions can only be approximated through realizations of $\bm{z}(x)$. A realization of the stochastic process, $z$, can be written as:
\begin{equation*}
z_t(x) = \widehat{a}(x) + \sum^{K}_{k=1}\widehat{\beta}_{t,k}\widehat{\phi}_k(x)+e_t(x),\qquad t=1,2,\dots,n,
\end{equation*}
where $\widehat{\beta}_{t,k}$ is the $k$\textsuperscript{th} estimated score for the $t$\textsuperscript{th} year and $e_t(x)$ denotes residual function.
 
\begin{figure}[!htbp]
\centering
\subfloat[First dynamic principal component using the raw data]
{\includegraphics[width=8cm]{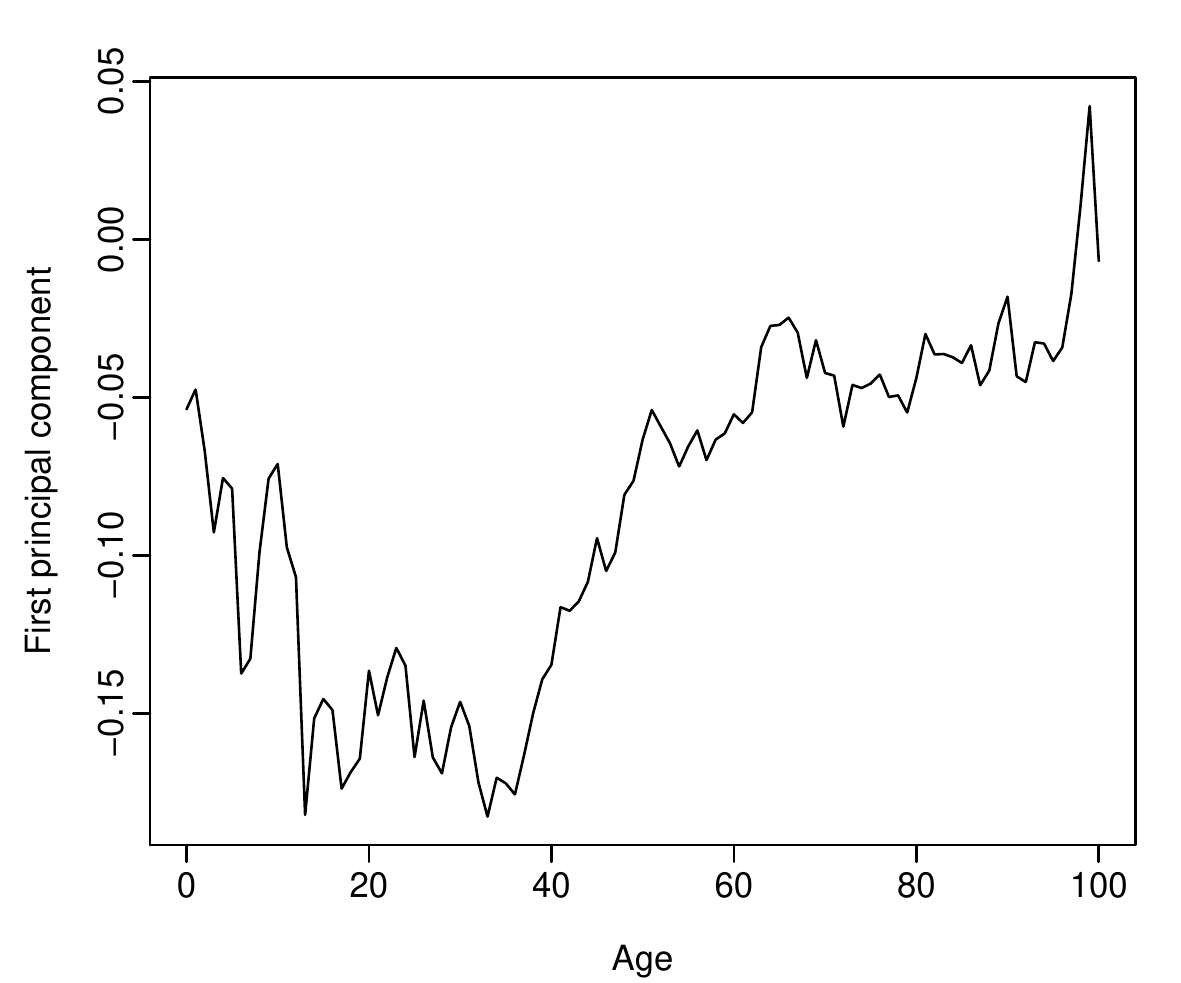}}
\qquad
\subfloat[First static principal component using the raw data]
{\includegraphics[width=8cm]{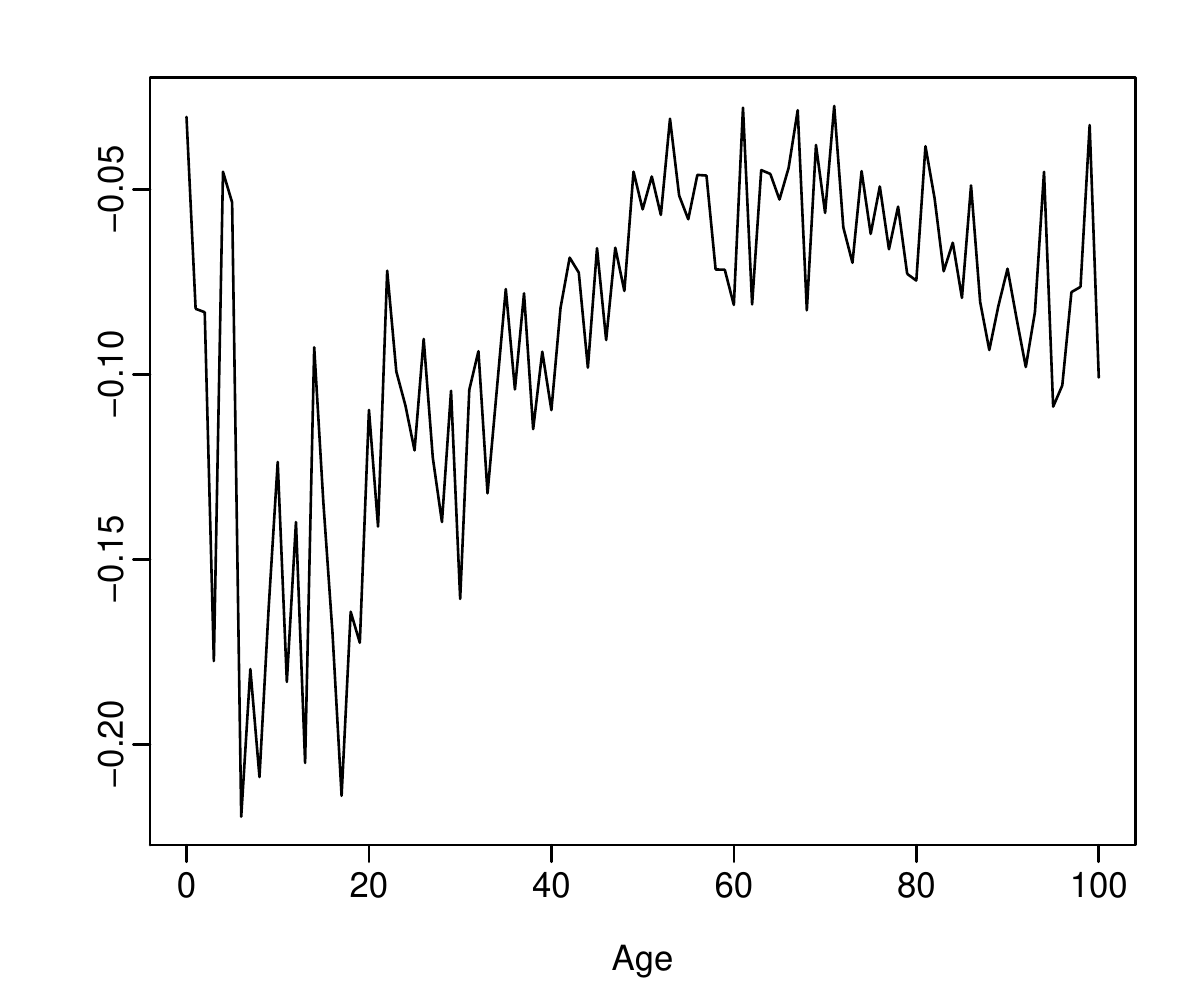}}
\\
\subfloat[First dynamic principal component using the smooth data]
{\includegraphics[width=8cm]{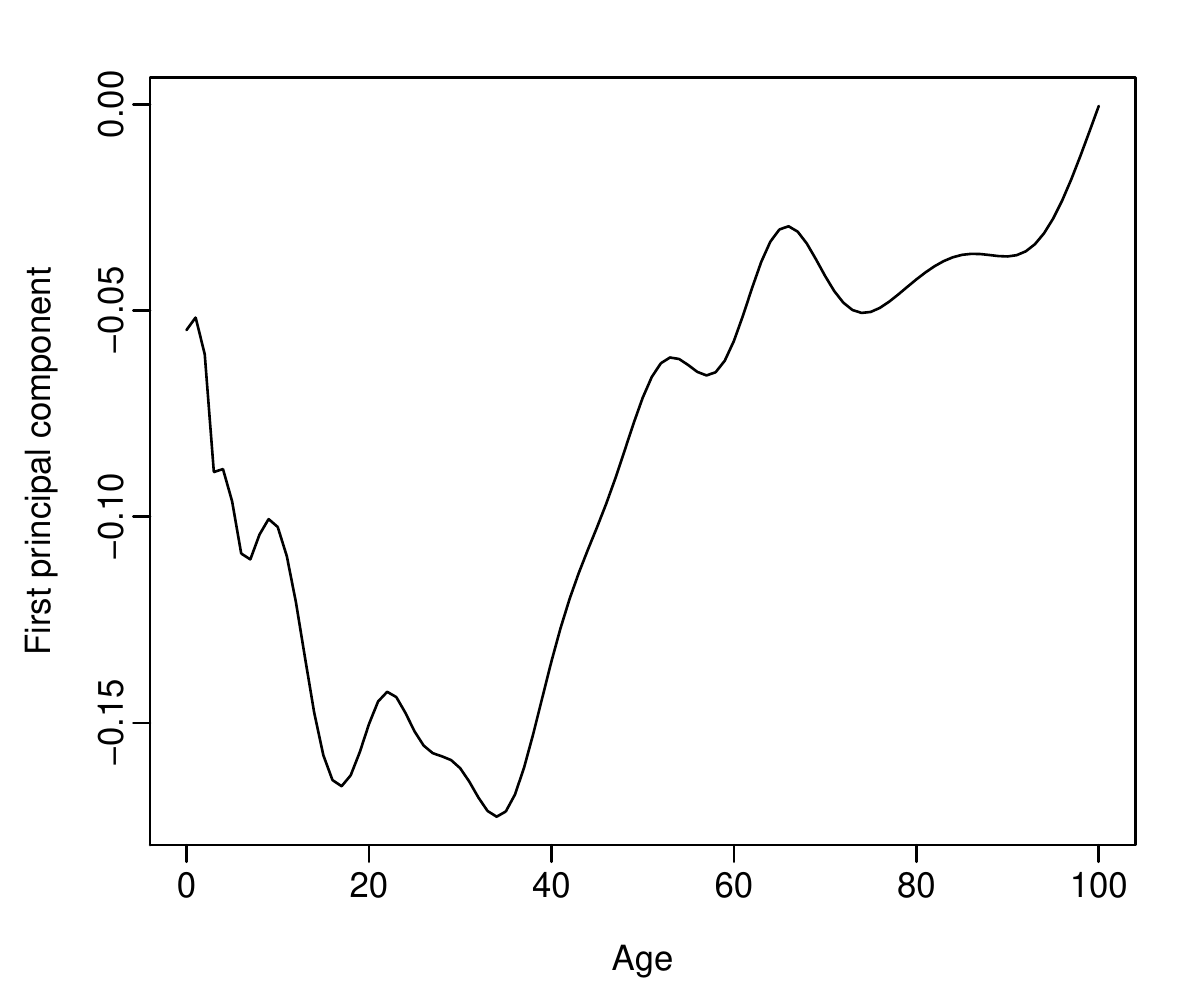}}
\qquad
\subfloat[First static principal component using the smooth data]
{\includegraphics[width=8cm]{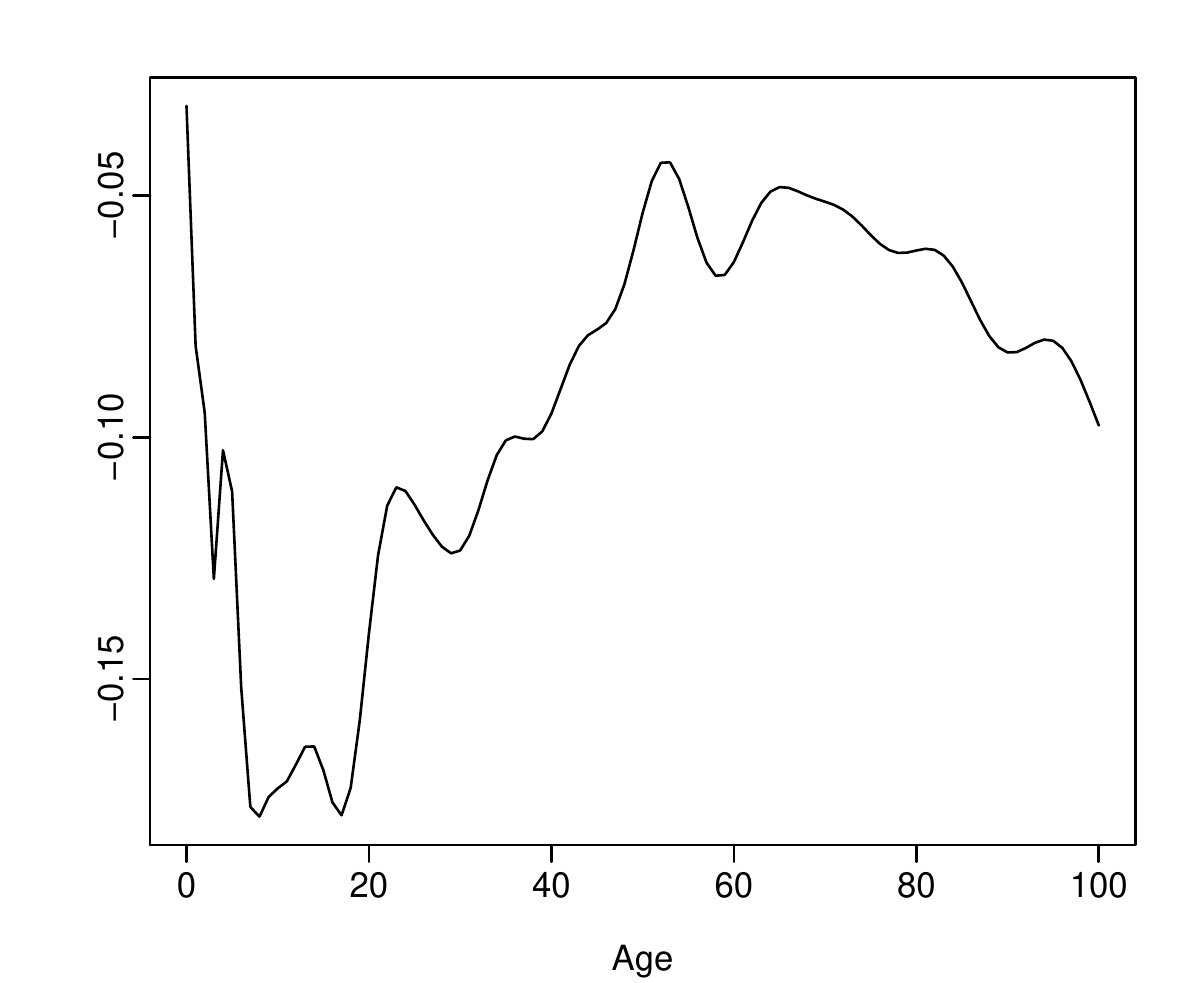}}
\caption{\small First eigenfunction extracted from the static principal component decomposition based on the sample variance, and first eigenfunction extracted from the dynamic principal component decomposition based on the sample long-run covariance. We considered the raw US mortality data in the top row, as well as the smooth US mortality data in the bottom row.}\label{fig:eigen_fun}
\end{figure}

In Figure~\ref{fig:eigen_fun}, we present the first eigenfunction extracted from the sample variance and sample long-run covariance, respectively for the US female mortality. Visually, the first dynamic principal component appears differently to the first static principal component. Conditioning on the estimated mean function $\widehat{a}(x)$, the estimated dynamic functional principal components $\Psi = \{\phi_1, \dots, \phi_K\}$, and observed mortality rate improvement, $\bm{z}(x)$, the $h$-step-ahead point forecast of $z_{n+h}(x)$ is
\begin{align*}
\widehat{z}_{n+h|n}(x) &= \text{E}\left[z_{n+h}(x)|\bm{z}(x), \widehat{a}(x), \bm{\Psi}\right] \\
&= \widehat{a}(x) + \sum^K_{k=1}\widehat{\beta}_{n+h|n,k}\widehat{\phi}_k(x),
\end{align*}
where $\widehat{\phi}_k(x)$ denotes the $k$\textsuperscript{th} estimated functional principal component, $\widehat{\beta}_{n+h|n,k}$ denotes the $k$\textsuperscript{th} estimated principal component scores obtained via a univariate forecasting method, and $K < n$ denotes the number of principal components retained. In practice, the optimal value of $K$ can be selected by explaining at least 85\% of total variation in the data, refer to Eq.~\eqref{eq:ncomp}.

\subsection{Constructing prediction intervals}

We considered a nonparametric bootstrap method for constructing prediction intervals \citep[for details, see][]{HS09}. The source of uncertainty stemmed from the estimation error in the principal component scores and model residual errors.

Using a univariate time series forecasting method, we could obtain multi-step-ahead forecasts for the principal component scores, $\{\widehat{\beta}_{1,k},\dots,\widehat{\beta}_{n,k}\}$. Let the $h$-step-ahead forecast errors denote:
\begin{equation*}
\widehat{\xi}_{t, h, k} = \widehat{\beta}_{t,k} - \widehat{\beta}_{t|t-h,k}, \qquad t=h+1,\dots,n.
\end{equation*}
The estimation errors were then sampled with replacement to give a bootstrap sample of $\beta_{n+h,k}$:
\begin{equation*}
\widehat{\beta}_{n+h|n,k}^{(b)} = \widehat{\beta}_{n+h|n,k} + \widehat{\xi}_{\ast, h, k}^{(b)}, \qquad b=1,\dots,B,
\end{equation*}
where $\widehat{\xi}_{\ast, h, k}^{(b)}$ are sampled with replacement from $\widehat{\xi}_{t, h, k}$, and $B$ denotes the number of bootstrap replications. As long as the first $K$ principal components approximate the data relatively well, the model residual should be random noise. We could bootstrap the model error by sampling with replacement from the residual term $\{\widehat{e}_1(x),\dots,\widehat{e}_{n}(x)\}$.

Through combining the two sources of uncertainty, we obtained $B$ variants for $\widehat{z}^{(b)}_{n+h|n}(x)$:
\begin{equation*}
\widehat{z}_{n+h|n}^{(b)}(x) = \widehat{a}(x) + \sum^{K}_{k=1}\widehat{\beta}_{n+h|n,k}^{(b)}\widehat{\phi}_k(x) + \widehat{e}_{n+h|n}^{(b)}(x).
\end{equation*}
Pointwise prediction intervals were produced from the bootstrap variants using quantiles.

\section{Results}\label{sec:results}
 
\subsection{Forecast evaluation}

We presented 24 countries with data that began in 1950 and ended in the final year listed in Table~\ref{tab:1}. We retained the final 30 observations for forecasting evaluation, while the remaining observations were treated as initial fitting observations, from which we produced the \textit{one-step-ahead} forecast (i.e., one-year-ahead forecast). Via an expanding window approach, we re-estimated the parameters in the time-series forecasting models by increasing the fitted observations by one year and producing the one-step-ahead forecast. We iterated this process by increasing the sample size by one year until reaching the end of the data period. This process produced 30 one-step-ahead forecasts. We compared these forecasts with the holdout samples to determine the out-of-sample forecast accuracy. 

\subsection{Forecast error criteria}

To evaluate the point forecast accuracy, we considered the mean absolute forecast error (MAFE) and root mean squared forecast error (RMSFE). These criteria measured the closeness of the forecasts in comparison with the actual values of the variable being forecast, regardless of the direction of forecast errors. The MAFE and RMSFE are defined as:
\begin{align*}
\text{MAFE}_h &= \frac{1}{p\times q} \sum^{q}_{j=1}\sum^{p}_{i=1}\left|m_{j}(x_i)-\widehat{m}_{j|j-h}(x_i)\right|,\\
\text{RMSFE}_h &= \frac{1}{p\times q} \sum^{q}_{j=1}\sum^{p}_{i=1}\sqrt{\left[m_{j}(x_i)-\widehat{m}_{j|j-h}(x_i)\right]^2},
\end{align*}
where $q$ represents the number of years in the forecasting period, $p\times q$ counts the total number of data points in the forecasting period, $m_{j}(x_i)$ represents the actual holdout sample for age $x_i$ in year $j$, and $\widehat{m}_{j}(x_i)$ represents the forecasts for the holdout sample. 

To evaluate the pointwise interval forecast accuracy, we considered the coverage probability deviance (CPD) of \cite{Shang12} and interval score criterion of \cite{GR07}. We considered the common case of the symmetric $100(1-\alpha)\%$ prediction intervals, with lower and upper bounds that were predictive quantiles at $\alpha/2$ and $1-\alpha/2$, denoted by $\widehat{m}_{j}^l(x_i)$ and $\widehat{m}_{j}^u(x_i)$. The CPD allows comparison of interval forecast accuracy for each method by measuring the differences between the empirical coverage and nominal coverage probabilities. The CPD is defined as
\begin{equation*}
\left|\frac{\mathds{1}\left\{m_{j}(x_i)<\widehat{m}_{j}^l(x_i)\right\} + \mathds{1}\left\{m_{j}(x_i)>\widehat{m}_{j}^u(x_i)\right\}}{p\times q} - \alpha\right|,
\end{equation*}
where $\mathds{1}\{\cdot\}$ denotes binary indicator, and $\alpha$ denotes the level of significance, customarily $\alpha = 0.2$. 

As defined by \cite{GR07}, a scoring rule for evaluating the pointwise interval forecast accuracy at time point $x_i$ is
\begin{align*}
S_{\alpha}[\widehat{m}_{j}^l(x_i), \widehat{m}_{j}^u(x_i); m_{j}(x_i)] = \left[\widehat{m}_{j}^u(x_i) - \widehat{m}_{j}^l(x_i)\right] &+ \frac{2}{\alpha}\left[\widehat{m}_{j}^l(x_i) - m_{j}(x_i)\right]\mathds{1}\left\{m_{j}(x_i)<\widehat{m}_{j}^l(x_i)\right\} \\
&+\frac{2}{\alpha}\left[m_{j}(x_i) - \widehat{m}_{j}^u(x_i)\right]\mathds{1}\left\{m_{j}(x_i)>\widehat{m}_{j}^u(x_i)\right\},
\end{align*}
The optimal interval score is achieved when $m_{j}(x_i)$ lies between $\widehat{m}^l_{j}(x_i)$ and $\widehat{m}^u_{j}(x_i)$, with the distance between the upper bound and lower bound being minimal. To obtain summary statistics of the interval score, we take the mean interval score across different ages and forecasting years. The mean interval score can be expressed as
\begin{equation*}
\overline{S}_{\alpha, h} = \frac{1}{p\times q}\sum^q_{j=1}\sum^p_{i=1}S_{\alpha}[\widehat{m}_{j}^l(x_i), \widehat{m}_{j}^u(x_i); m_{j}(x_i)].
\end{equation*}

\subsection{Comparisons of forecast errors}

\begin{figure}[!htbp]
\centering
{\includegraphics[width=8.7cm]{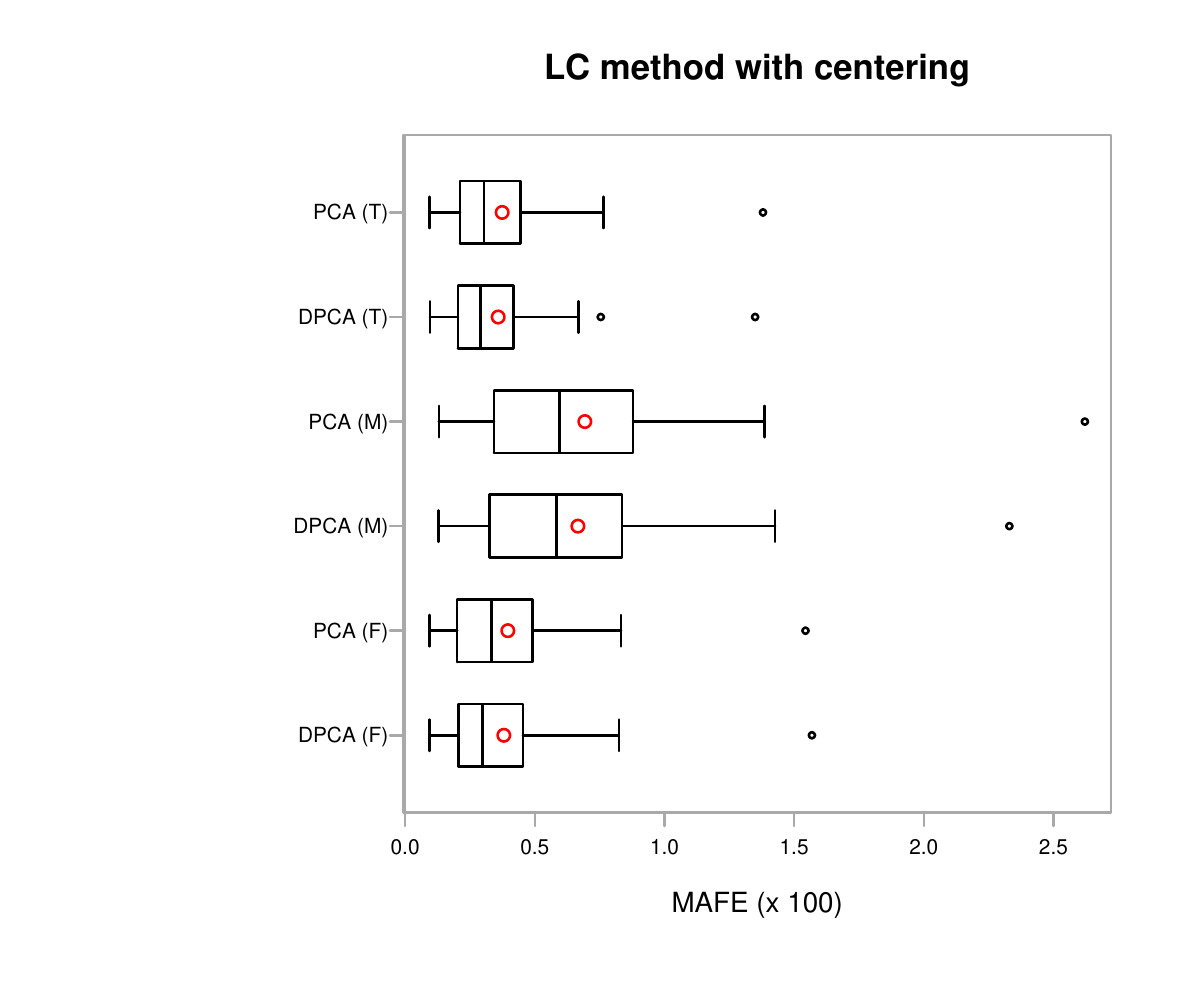}}
\quad
{\includegraphics[width=8.7cm]{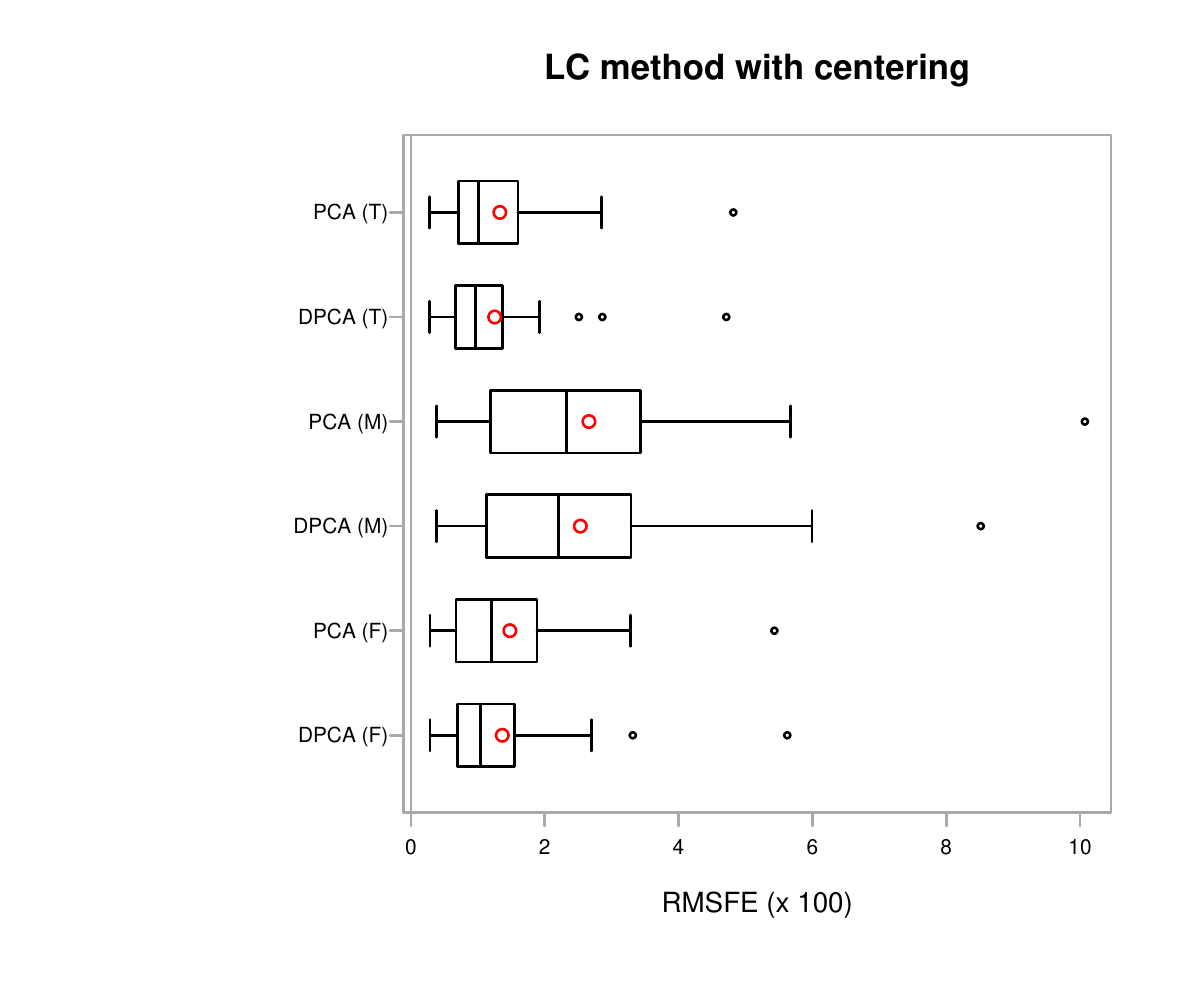}}
\\
{\includegraphics[width=8.7cm]{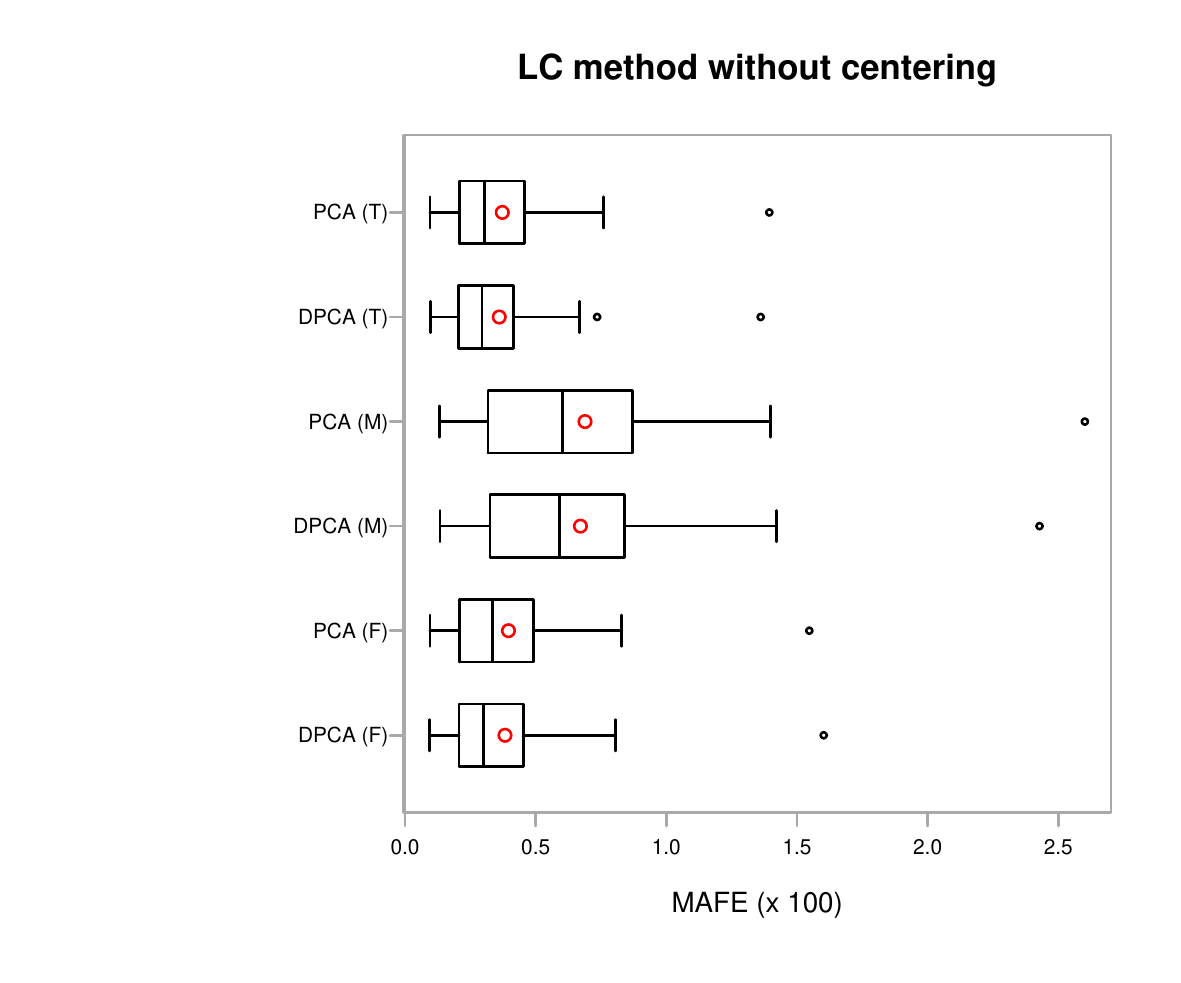}}
\quad
{\includegraphics[width=8.7cm]{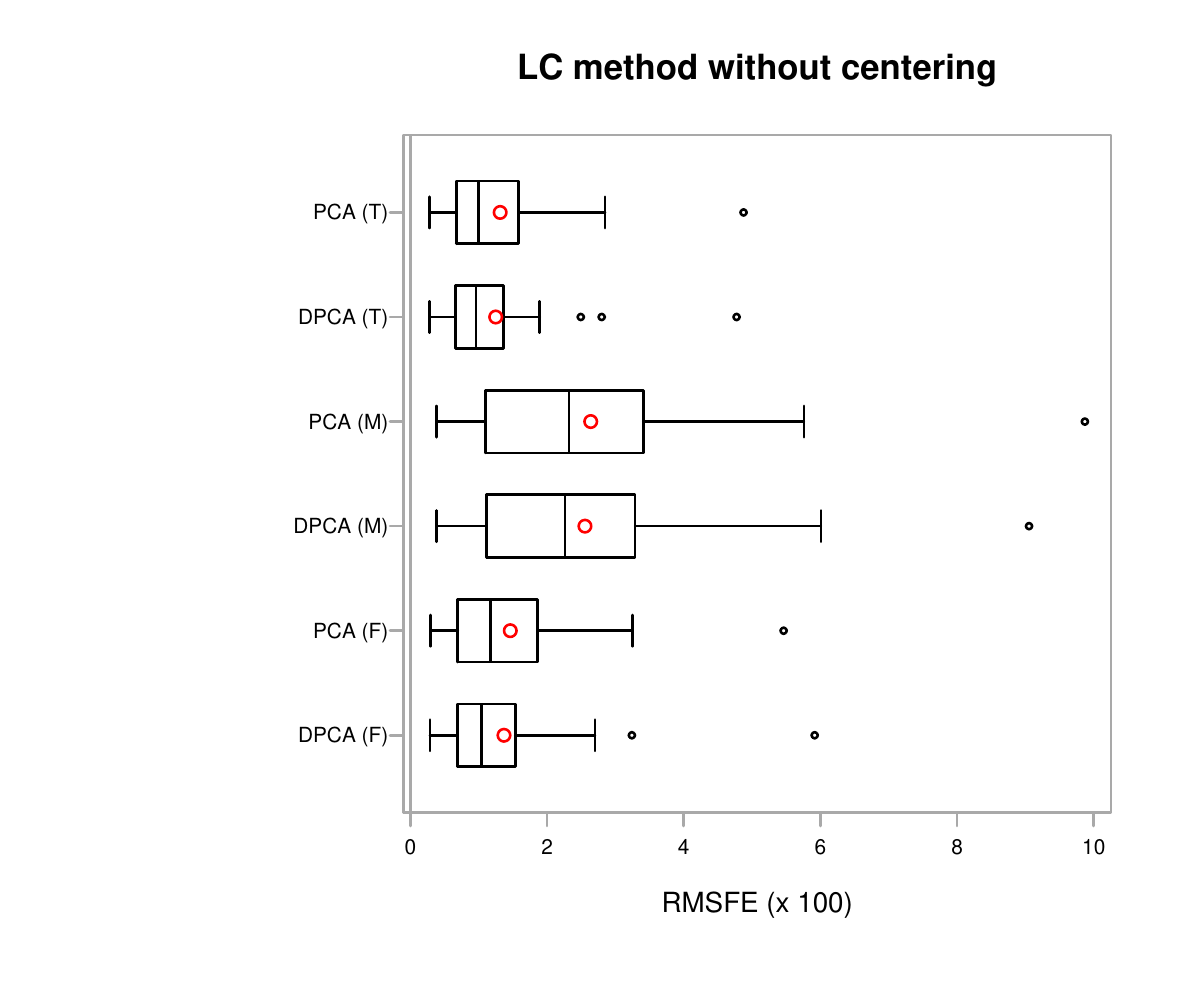}}
\\ 
{\includegraphics[width=8.7cm]{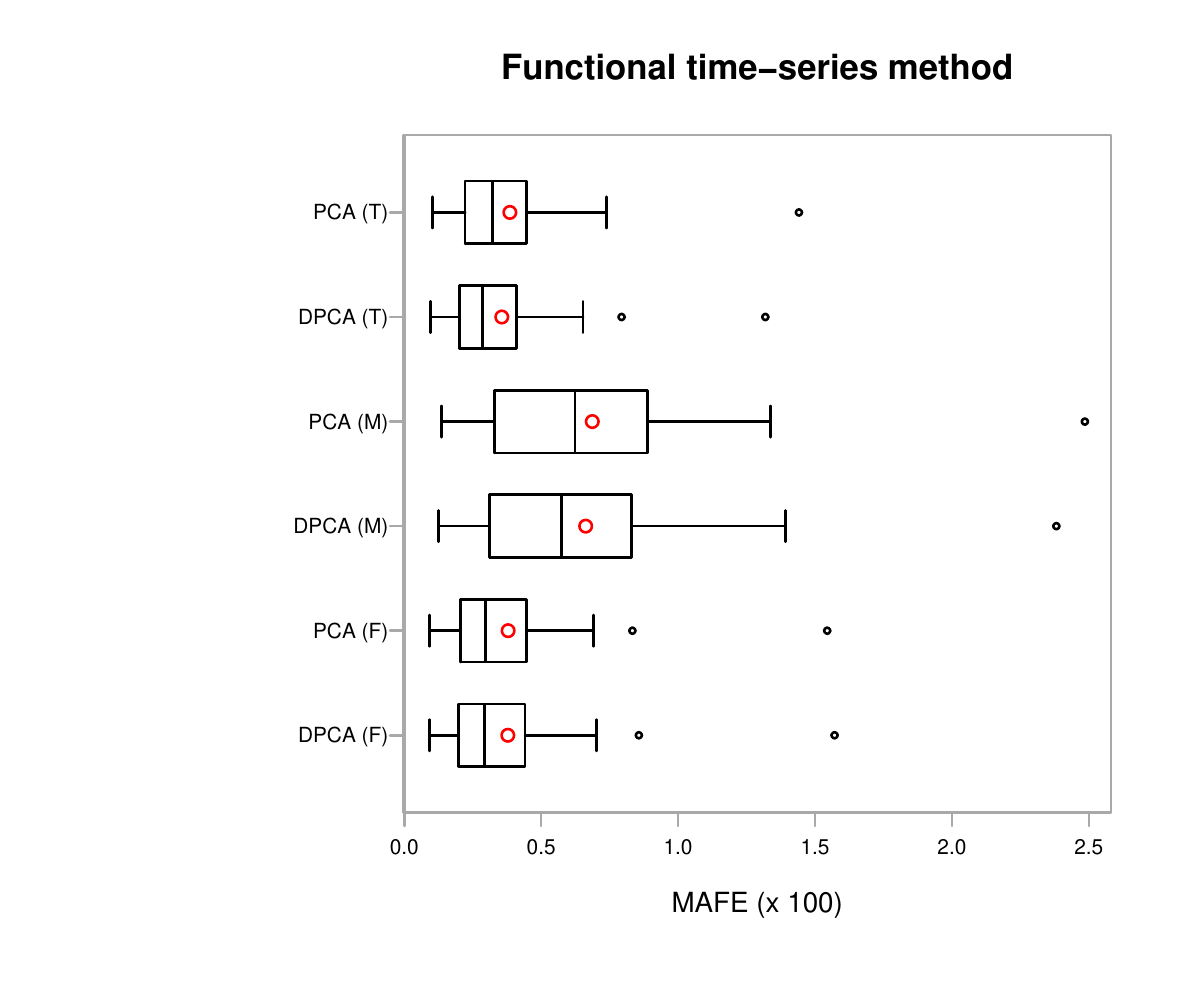}}
\quad
{\includegraphics[width=8.7cm]{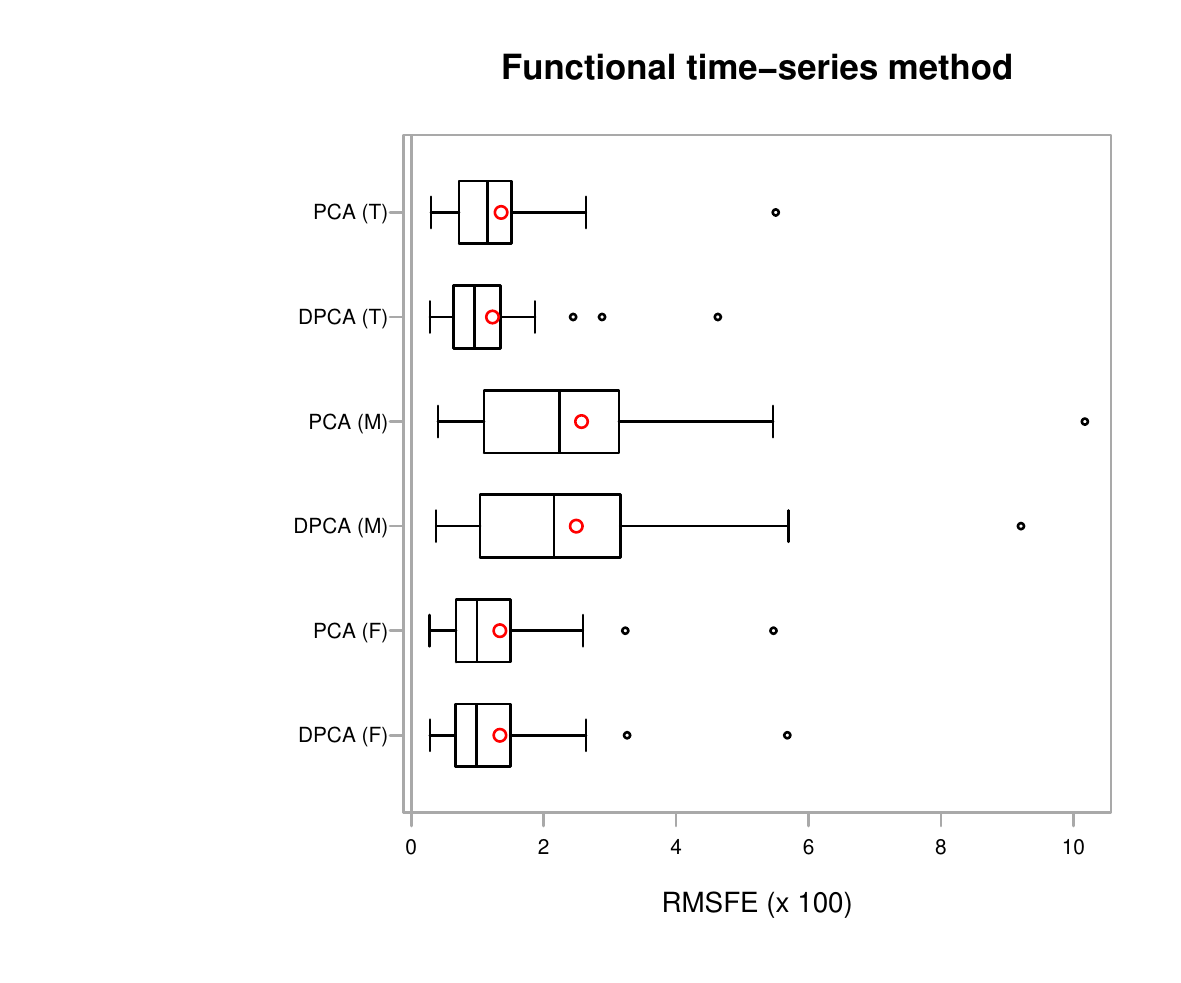}}
\caption{\small Boxplots of one-step-ahead point forecast errors between the DPCA and static PCA using the LC method with and without centering and functional time-series method. The red circle represents the mean.}\label{fig:point}
\end{figure}

In Figure~\ref{fig:point}, we compare the one-step-ahead point forecast errors between the dynamic and static principal component regression models using the LC method with and without centering and functional time-series method. 

\begin{figure}[!htbp]
\centering
{\includegraphics[width=8.7cm]{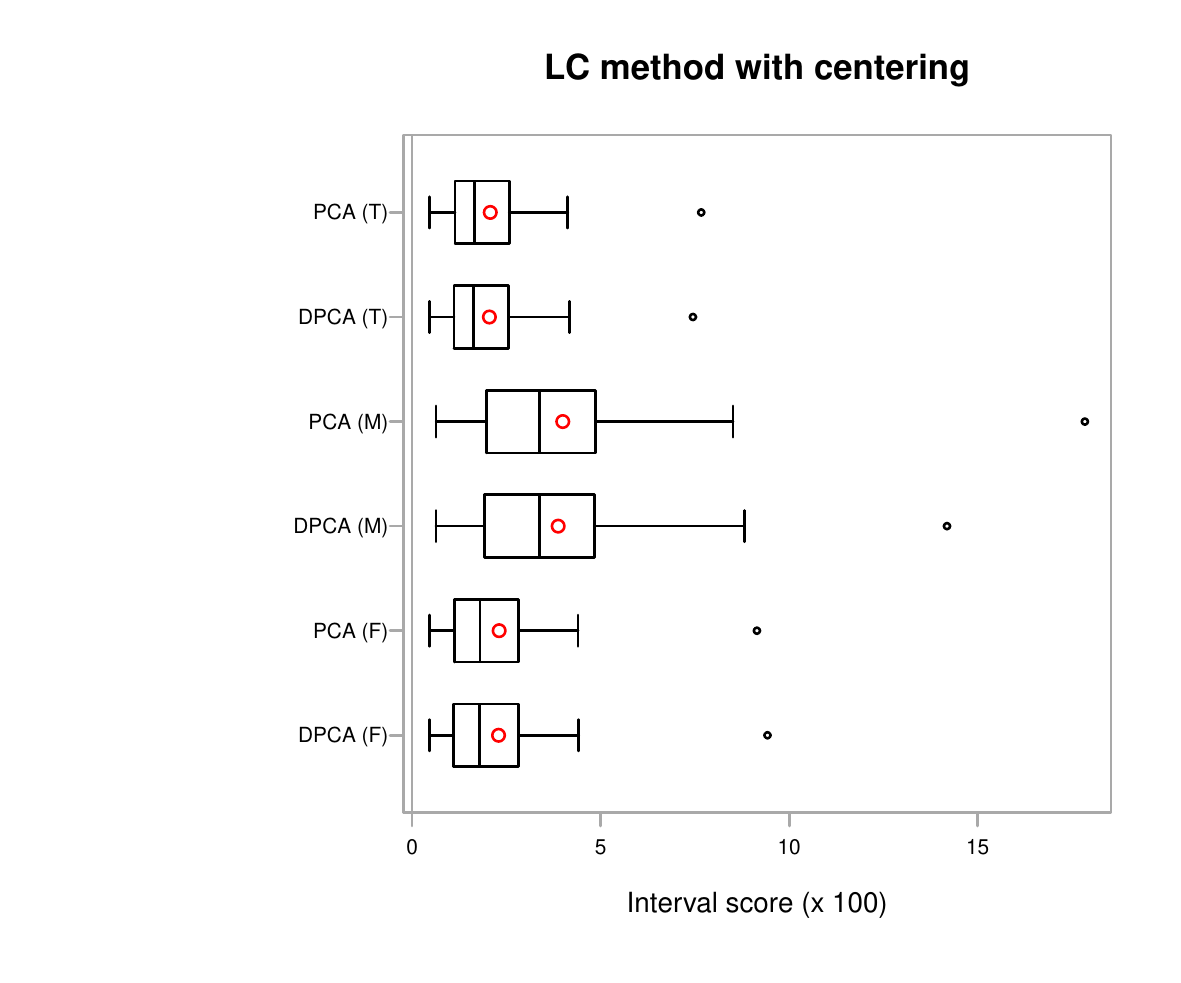}}
\quad
{\includegraphics[width=8.7cm]{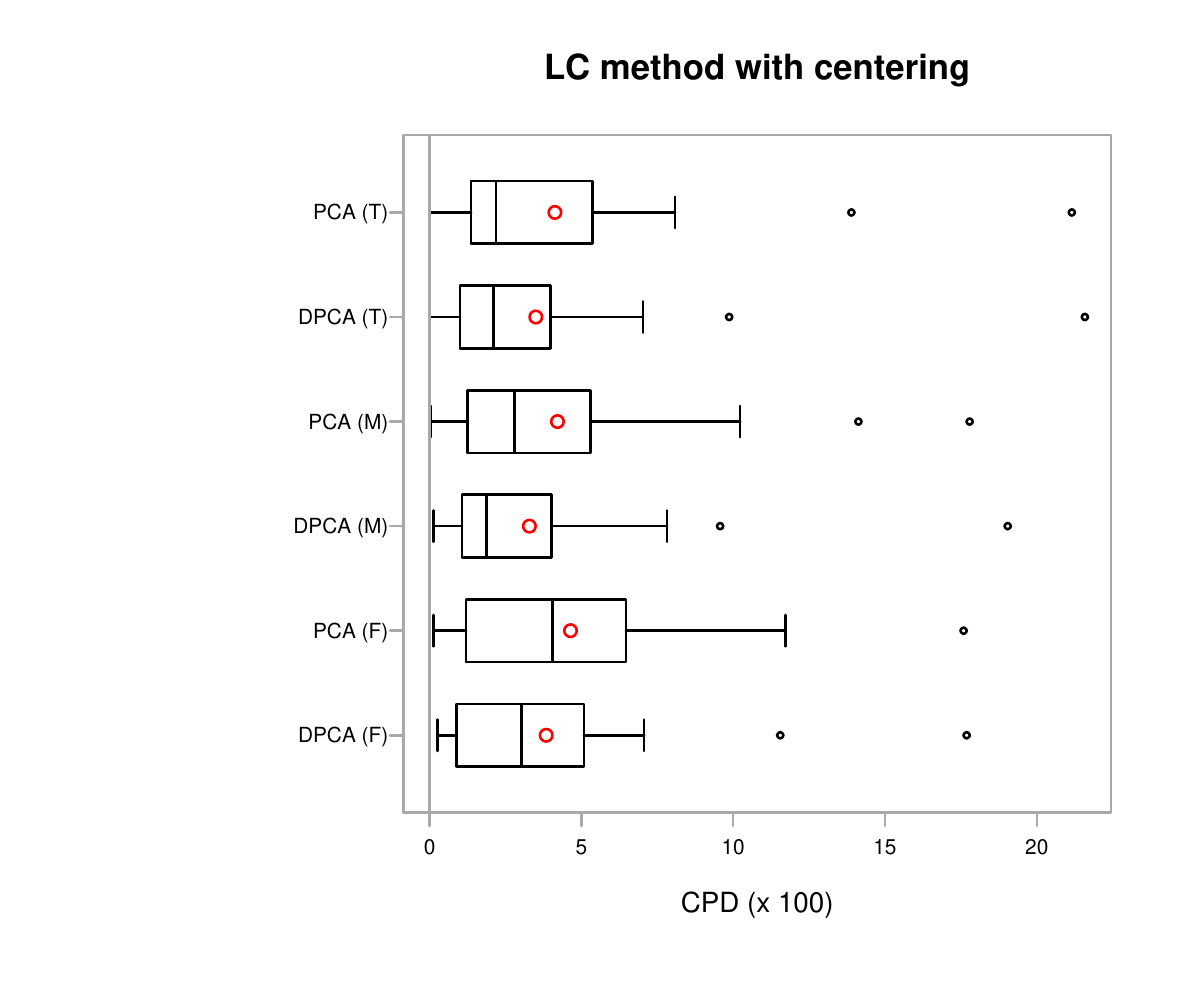}}
\\
\vspace{.2in}
{\includegraphics[width=8.7cm]{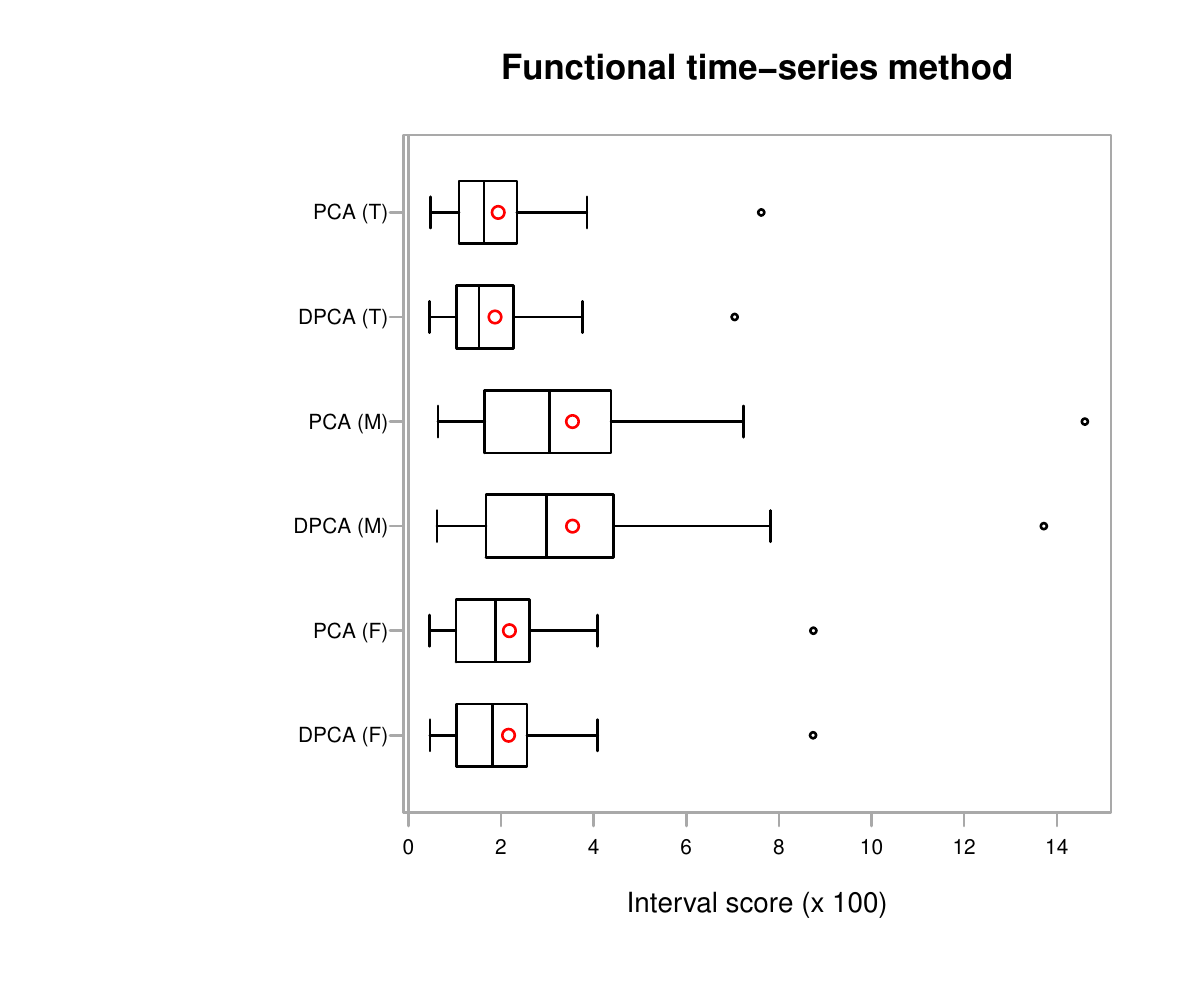}}
\quad
{\includegraphics[width=8.7cm]{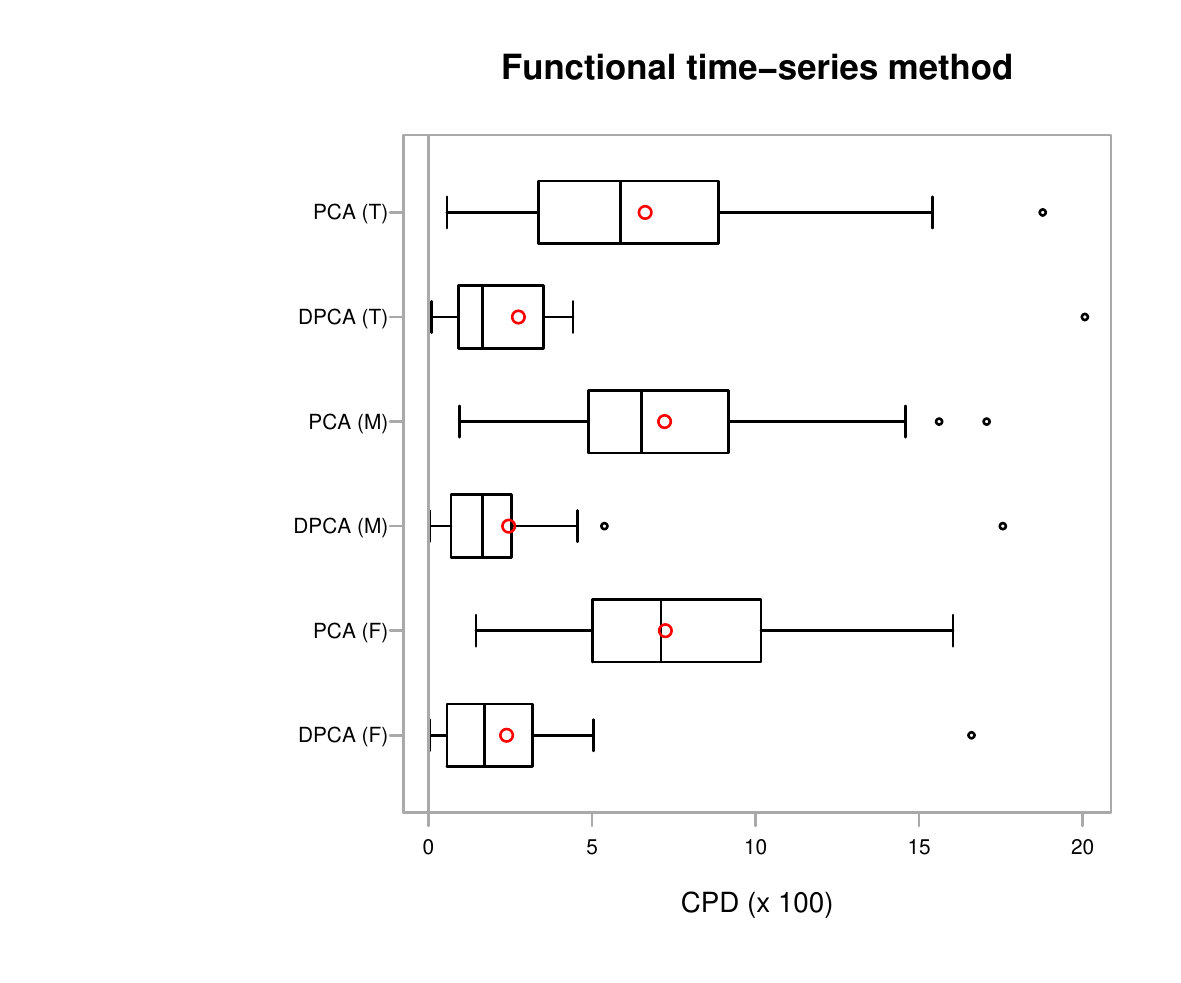}}
\caption{\small Boxplots of one-step-ahead interval forecast errors between the dynamic and static principal component regression using the LC method with centering and functional time-series method. The red circle represents the mean.}\label{fig:interval}
\end{figure}

In Figure~\ref{fig:interval}, we compare the one-step-ahead interval forecast errors between the dynamic and static principal component regression models using the LC method with centering and functional time-series method.

In Table~\ref{tab:point_accuracy}, we compare the summary statistics of the one-step-ahead point forecast errors between the dynamic and static principal component regression models using the LC method with and without centering and functional time-series method. Given that the point forecast errors between the LC method with and without centering are marginal, in Table~\ref{tab:interval_accuracy}, we compare the summary statistics of the one-step-ahead interval forecast accuracy between the dynamic and static principal component regression models using the LC method with centering and functional time-series method.

\begin{center}
\begin{small}
\begin{longtable}{@{}lrrrrrrrrrrrr@{}}
\caption{One-step-ahead point forecast errors between the DPCA and static PCA using the LC method with and without centering and functional time-series method. Due to limitations of space, we report the summary statistics of the forecast errors obtained from the 24 countries.}\label{tab:point_accuracy}\\
\toprule
	& \multicolumn{6}{c}{MAFE $\times 100$} & \multicolumn{6}{c}{RMSFE $\times 100$} \\
  		& \multicolumn{2}{c}{Female} & \multicolumn{2}{c}{Male} & \multicolumn{2}{c}{Total} & \multicolumn{2}{c}{Female} & \multicolumn{2}{c}{Male} & \multicolumn{2}{c}{Total} \\
 Statistic & DPCA & PCA & DPCA & PCA & DPCA & PCA & DPCA & PCA & DPCA & PCA & DPCA & PCA \\\midrule
\endfirsthead
	& \multicolumn{6}{c}{MAFE $\times 100$} & \multicolumn{6}{c}{RMSFE $\times 100$} \\
  		& \multicolumn{2}{c}{Female} & \multicolumn{2}{c}{Male} & \multicolumn{2}{c}{Total} & \multicolumn{2}{c}{Female} & \multicolumn{2}{c}{Male} & \multicolumn{2}{c}{Total} \\
 Statistic & DPCA & PCA & DPCA & PCA & DPCA & PCA & DPCA & PCA & DPCA & PCA & DPCA & PCA \\ 
\endhead
\multicolumn{13}{r}{{Continued on next page}} \\ \hline
\endfoot
\endlastfoot
\multicolumn{6}{l}{\hspace{-.08in}{LC method with centering}} \\
Min. & 0.094 & 0.094 & 0.129 & 0.130 & 0.095 & 0.095 & 0.288 & 0.288 & 0.382 & 0.384 & 0.280 & 0.279 \\  
  1st Qu. & 0.214 & 0.208 & 0.339 & 0.357 & 0.211 & 0.217 & 0.720 & 0.699 & 1.154 & 1.247 & 0.687 & 0.724 \\ 
  Median & 0.299 & 0.332 & 0.583 & 0.596 & 0.291 & 0.304 & 1.039 & 1.204 & 2.205 & 2.322 & 0.967 & 1.011 \\  
  Mean & 0.381 & 0.396 & 0.666 & 0.693 & 0.359 & 0.374 & 1.366 & 1.479 & 2.532 & 2.661 & 1.252 & 1.330 \\  
  3rd Qu. & 0.452 & 0.488 & 0.837 & 0.867 & 0.416 & 0.440 & 1.535 & 1.867 & 3.218 & 3.385 & 1.360 & 1.585 \\ 
  Max. & 1.569 & 1.544 & 2.330 & 2.621 & 1.349 & 1.380 & 5.625 & 5.433 & 8.516 & 10.072 & 4.714 & 4.819 \\  
\midrule
\multicolumn{6}{l}{\hspace{-.08in}{LC method without centering}} \\
Min. & 0.094 & 0.096 & 0.134 & 0.131 & 0.097 & 0.096 & 0.286 & 0.294 & 0.381 & 0.384 & 0.281 & 0.283 \\ 
  1st Qu. & 0.215 & 0.214 & 0.338 & 0.326 & 0.211 & 0.215 & 0.712 & 0.714 & 1.141 & 1.103 & 0.681 & 0.695 \\  
  Median & 0.300 & 0.335 & 0.590 & 0.602 & 0.294 & 0.305 & 1.036 & 1.173 & 2.263 & 2.318 & 0.959 & 0.997 \\  
  Mean & 0.383 & 0.396 & 0.671 & 0.689 & 0.361 & 0.372 & 1.370 & 1.462 & 2.555 & 2.639 & 1.249 & 1.314 \\  
  3rd Qu. & 0.453 & 0.488 & 0.837 & 0.857 & 0.415 & 0.449 & 1.524 & 1.855 & 3.217 & 3.375 & 1.350 & 1.526 \\  
  Max. & 1.602 & 1.547 & 2.428 & 2.602 & 1.361 & 1.394 & 5.916 & 5.462 & 9.053 & 9.870 & 4.774 & 4.875 \\ 
\midrule
\multicolumn{6}{l}{\hspace{-.08in}{Functional time-series method}} \\
Min. & 0.092 & 0.093 & 0.126 & 0.136 & 0.095 & 0.103 & 0.281 & 0.276 & 0.376 & 0.404 & 0.283 & 0.299 \\  
  1st Qu. & 0.210 & 0.210 & 0.319 & 0.341 & 0.208 & 0.233 & 0.702 & 0.687 & 1.042 & 1.108 & 0.667 & 0.748 \\  
  Median & 0.292 & 0.296 & 0.573 & 0.624 & 0.285 & 0.322 & 0.984 & 0.995 & 2.156 & 2.238 & 0.952 & 1.149 \\  
  Mean & 0.378 & 0.379 & 0.662 & 0.687 & 0.356 & 0.386 & 1.339 & 1.338 & 2.492 & 2.571 & 1.226 & 1.357 \\  
  3rd Qu. & 0.441 & 0.444 & 0.823 & 0.887 & 0.406 & 0.435 & 1.491 & 1.495 & 3.092 & 3.123 & 1.339 & 1.510 \\  
  Max. & 1.571 & 1.545 & 2.381 & 2.486 & 1.319 & 1.441 & 5.679 & 5.470 & 9.207 & 10.172 & 4.630 & 5.504 \\ 
      \bottomrule
\end{longtable}
\end{small}
\end{center}

\begin{center}
\begin{small}
\tabcolsep 0.072in
\begin{longtable}{@{}lrrrrrrrrrrrr@{}}
\caption{One-step-ahead interval forecast errors between the dynamic and static principal component regression using the LC method with centering and functional time-series method. Due to limitations of space, we report the summary statistics of the forecast errors obtained from the 24 countries.}\label{tab:interval_accuracy}\\
\toprule
	& \multicolumn{6}{c}{Score $\times 100$} & \multicolumn{6}{c}{CPD $\times 100$} \\
  		& \multicolumn{2}{c}{Female} & \multicolumn{2}{c}{Male} & \multicolumn{2}{c}{Total} & \multicolumn{2}{c}{Female} & \multicolumn{2}{c}{Male} & \multicolumn{2}{c}{Total} \\
 Statistic & DPCA & PCA & DPCA & PCA & DPCA & PCA & DPCA & PCA & DPCA & PCA & DPCA & PCA \\\midrule
\endfirsthead
	& \multicolumn{6}{c}{Score $\times 100$} & \multicolumn{6}{c}{CPD $\times 100$} \\
  		& \multicolumn{2}{c}{Female} & \multicolumn{2}{c}{Male} & \multicolumn{2}{c}{Total} & \multicolumn{2}{c}{Female} & \multicolumn{2}{c}{Male} & \multicolumn{2}{c}{Total} \\
 Statistic & DPCA & PCA & DPCA & PCA & DPCA & PCA & DPCA & PCA & DPCA & PCA & DPCA & PCA \\ 
\endhead
\midrule
\multicolumn{13}{r}{{Continued on next page}} \\ 
\endfoot
\endlastfoot
\multicolumn{6}{l}{\hspace{-.08in}{LC method with centering}} \\
 Min. & 0.461 & 0.463 & 0.629 & 0.637 & 0.466 & 0.466 & 0.264 & 0.132 & 0.132 & 0.033 & 0.000 & 0.000 \\ 
  1st Qu. & 1.148 & 1.150 & 1.931 & 1.967 & 1.137 & 1.152 & 0.891 & 1.312 & 1.114 & 1.271 & 1.048 & 1.411 \\  
  Median & 1.790 & 1.795 & 3.376 & 3.385 & 1.633 & 1.646 & 3.036 & 4.043 & 1.881 & 2.805 & 2.112 & 2.195 \\  
  Mean & 2.291 & 2.307 & 3.873 & 3.994 & 2.048 & 2.071 & 3.842 & 4.642 & 3.288 & 4.212 & 3.500 & 4.130 \\  
  3rd Qu. & 2.806 & 2.772 & 4.779 & 4.809 & 2.506 & 2.506 & 5.000 & 6.353 & 3.870 & 5.272 & 3.911 & 5.256 \\  
  Max. & 9.424 & 9.144 & 14.185 & 17.842 & 7.446 & 7.666 & 17.690 & 17.591 & 19.043 & 17.789 & 21.584 & 21.155 \\
\midrule
\multicolumn{6}{l}{\hspace{-.08in}{Functional time-series method}} \\
Min. & 0.463 & 0.458 & 0.615 & 0.640 & 0.459 & 0.477 & 0.033 & 1.452 & 0.033 & 0.957 & 0.099 & 0.561 \\  
  1st Qu. & 1.093 & 1.057 & 1.706 & 1.678 & 1.065 & 1.138 & 0.578 & 5.066 & 0.726 & 5.058 & 0.924 & 3.614 \\  
  Median & 1.813 & 1.887 & 2.982 & 3.043 & 1.529 & 1.631 & 1.716 & 7.112 & 1.650 & 6.518 & 1.650 & 5.875 \\  
  Mean & 2.163 & 2.182 & 3.545 & 3.542 & 1.871 & 1.940 & 2.389 & 7.241 & 2.452 & 7.218 & 2.749 & 6.624 \\  
  3rd Qu. & 2.547 & 2.592 & 4.352 & 4.294 & 2.258 & 2.328 & 3.127 & 10.132 & 2.442 & 9.125 & 3.507 & 8.804 \\  
  Max. & 8.738 & 8.745 & 13.720 & 14.607 & 7.046 & 7.620 & 16.601 & 16.040 & 17.558 & 17.063 & 20.066 & 18.779 \\ 
          \bottomrule
\end{longtable}
\end{small}
\end{center}

We observed the following evidence:
\begin{asparaenum}
\item[1)] Averaged over the 24 mainly developed countries in Table~\ref{tab:1}, the dynamic principal component regression outperformed the static principal component regression in terms of point and interval forecast accuracies, as measured by summary statistics of the MAFE and RMSFE, CPD and mean interval score criteria. 
\item[2)] In contrast to the results of the LC method, it was advantageous to smooth the data before computing the long-run covariance and variance, because this approach generally produced smaller point and interval forecast errors. 
\item[3)] Among the female, male, and total series, it was generally easier to forecast the total series as evident from smaller forecast errors, while it was harder to forecast the male series as evident from larger forecast errors. This could be because the total variation of the male series was larger than that of the total series.
\end{asparaenum}
The superiority of the dynamic principal component regression could be because it captures temporal dependence better than the static principal component regression.

We also considered the five-step-ahead and 10-step-ahead point forecast accuracy, and found a marginal difference between the dynamic and static approaches. The results are provided in Appendix C. In time-series forecasting, many time-series extrapolation models were no longer optimal when we projected long term. As the forecast horizon increased, the proposed DPCA reduced back to the static PCA used in the LC model. The difference between the two derived crucially from the criterion used to extract latent components. In the DPCA, the criterion was long-run covariance, which was a sum of the variance and autocovariance. In the static PCA, the criterion was variance alone. In the long-term forecast, the distant future value was almost IID to the most recent value. In turn, the autocovariance was small, if it existed at all, at a long-term forecast horizon. Therefore, the proposed DPCA (almost) reduced back to the static PCA, and the DPCA did not display advantages over the static PCA at a long-term forecast horizon. In addition, as the forecast horizon increased, the forecast of principal component scores was likely to be centered around zero. When that occurred, the forecasts of mortality rates were not so informative and would likely center around the mean function.

\section{Discussion}\label{sec:5}

PCA performs dimension reduction (also known as data coarsening) with a minimal loss of information, and is a workhorse in time-series modeling of age-specific mortality data and application to annuity pricing in actuarial science. The core techniques in the existing demographic literature use static PCA and may not incorporate temporal covariance into the eigen-decomposition. When the temporal dependence is moderate or strong, the static principal components extracted from variance are no longer optimal. As an alternative, we proposed a DPCA where the principal components could be extracted from an estimated long-run covariance. 

The long-run covariance encompasses the autocovariance at lag 0 (i.e., variance), as well as the autocovariance at other lags. To estimate the long-run covariance, we considered a kernel sandwich estimator used in \cite{Andrews91}. A crucial parameter in the kernel sandwich estimator is the selection of optimal bandwidth. To determine the optimal lags, we presented a plug-in algorithm of \cite{RS16} to determine the optimal bandwidth parameter in the kernel sandwich estimator. 

Given that the estimation of the long-run covariance requires stationarity, we chose to work with mortality rate improvement via the forward transformation. Through using static PCA or DPCA, we modeled and forecast mortality rate improvement. Through backward transformation, we obtained the forecast mortality rate in the original scale. Using 24 mainly developed countries, we demonstrated improvement of point and interval forecast accuracies that the dynamic principal component regression entails when compared with the static analysis using the LC and functional time-series methods.

It is noteworthy that, given a sufficient number of static principal components, they can capture a similar amount of information that the dynamic approach entails. For some countries, the temporal dependency is very weak, and then the long-run covariance almost reduces to the variance alone. In that case, the static and dynamic approaches lead to the same or similar point and interval forecast accuracies. We observed that there are small differences between the static and dynamic principal component regression models for the five-step-ahead and 10-step-ahead forecasts. For the one-step-ahead forecasts, the differences in point and interval forecast accuracies were rather apparent. Thus, it should be recommended as a valuable technique in statistical modeling and short-term forecasting.

There are a number of ways in which the current paper could be further extended, and we briefly mention three. First, a future extension could be to take a cohort perspective on the mortality improvement rate, as taken by \cite{HR13}. Second, given that most, if not all, extrapolation methods do not perform well in long-term forecasts, it may be useful to propose a Bayesian version of dynamic functional PCA, where the prior knowledge can be incorporated. Finally, we have demonstrated the usefulness of the DPCA using the LC and functional time-series models; however, it can be applied to other mortality models that use PCA in whole or in part \citep[see, e.g.,][]{SBH11}.

\section*{Appendix A}

\subsection*{Estimation of long-run covariance}\label{sec:2.1}

Under the asymptotic mean squared normed error, \cite{RS16} show that the optimal bandwidth parameter $h_{\text{opt}}$ has the following forms:
\begin{align}
h_{\text{opt}}&=c_0n^{\frac{1}{1+2q}} \notag\\
c_0 &= \left(2q\|C^{(q)}\|^2\right)^{\frac{1}{1+2q}}\left\{\left[\|C\|^2+\Big(\int^1_0 C(u,u)du\Big)^2\right]\int^{\infty}_{-\infty}W_q^2(x)dx\right\}^{-\frac{1}{1+2q}}, \tag{A.1} \label{eq:c_0}
\end{align}
where $q$ denotes the order of derivative, and $W_q(x)$ denotes a kernel (weight) function of order $q$. The crux of the problem is that the quantities involving $C^{(q)}$ and $C$ in Eq.~\eqref{eq:c_0} are unknown, and we use a plug-in algorithm to estimate them, from which we obtain $\widehat{c}_0$ and $\widehat{h}_{\text{opt}}$.

The plug-in bandwidth selection method is given as follows:
\begin{enumerate}
\item[1)] Compute pilot estimates of $C^{(p)}$, for $p=0, q$:
\begin{equation*}
\widehat{C}_{h_1,q_1}^{(p)}(x,u) = \sum^{\infty}_{\ell = -\infty}W_{q_1}\left(\frac{\ell}{h_1}\right)|\ell|^p\widehat{\gamma}_l(x,u),
\end{equation*}
which utilizes an initial bandwidth choice $h_1 = h_1(n)$, and an initial kernel function $W_{q_1}$ of order $q_1$.
\item[2)] Estimate $c_0$ by
\begin{equation*}
\widehat{c}_0(h_1,q_1) = \left(2q\left\|\widehat{C}_{h_1,q_1}^{(q)}\right\|^2\right)^{\frac{1}{1+2q}}\left\{\left[\left\|C\right\|^2+\Big(\int^1_0C(u,u)du\Big)^2\right]\int^{\infty}_{-\infty}W_q^2(x)dx\right\}^{-\frac{1}{1+2q}},
\end{equation*}
where $W_q$ denotes a final kernel function of order $q$.
\item[3)] Use the bandwidth
\begin{equation*}
\widehat{h}_{\text{opt}}(h_1,q_1) = \widehat{c}_0(h_1,q_1)n^{\frac{1}{1+2q}}
\end{equation*}
in the definition of $\widehat{C}_{h,q}$ in Eq (5) in the manuscript.
\end{enumerate}

In terms of the initial and final weight functions, \cite{RS16} advocated the use of a flat-top weight function $W_{\infty}$ for the initial kernel function, and a Bartlett kernel function as the final kernel function. A flat-top weight function $W_{\infty}$ is of the form
\[ W_{\infty}(t) = \left\{ \begin{array}{ll}
         1 & \mbox{\qquad $0\leq |t|< k_1$};\\
        \frac{k_2-|t|}{k_2-k_1} & \mbox{\qquad $k_1\leq |t|\leq k_2$}; \\
        0 & \mbox{\qquad $|t|\geq k_2$}.\end{array} \right. \] 
where $k_2>k_1$. Let us take $k_2 = 1$ and $k_1=0.5$. The Bartlett weight function $W_1$ is of the form
\[ W_1 = \left\{ \begin{array}{ll}
         1-|x| & \mbox{\qquad for $|x| \leq 1$};\\
        0 & \mbox{\qquad otherwise}.\end{array} \right. \] 

\section*{Appendix B}

\begin{figure}[!htbp]
\centering
\includegraphics[width=\textwidth]{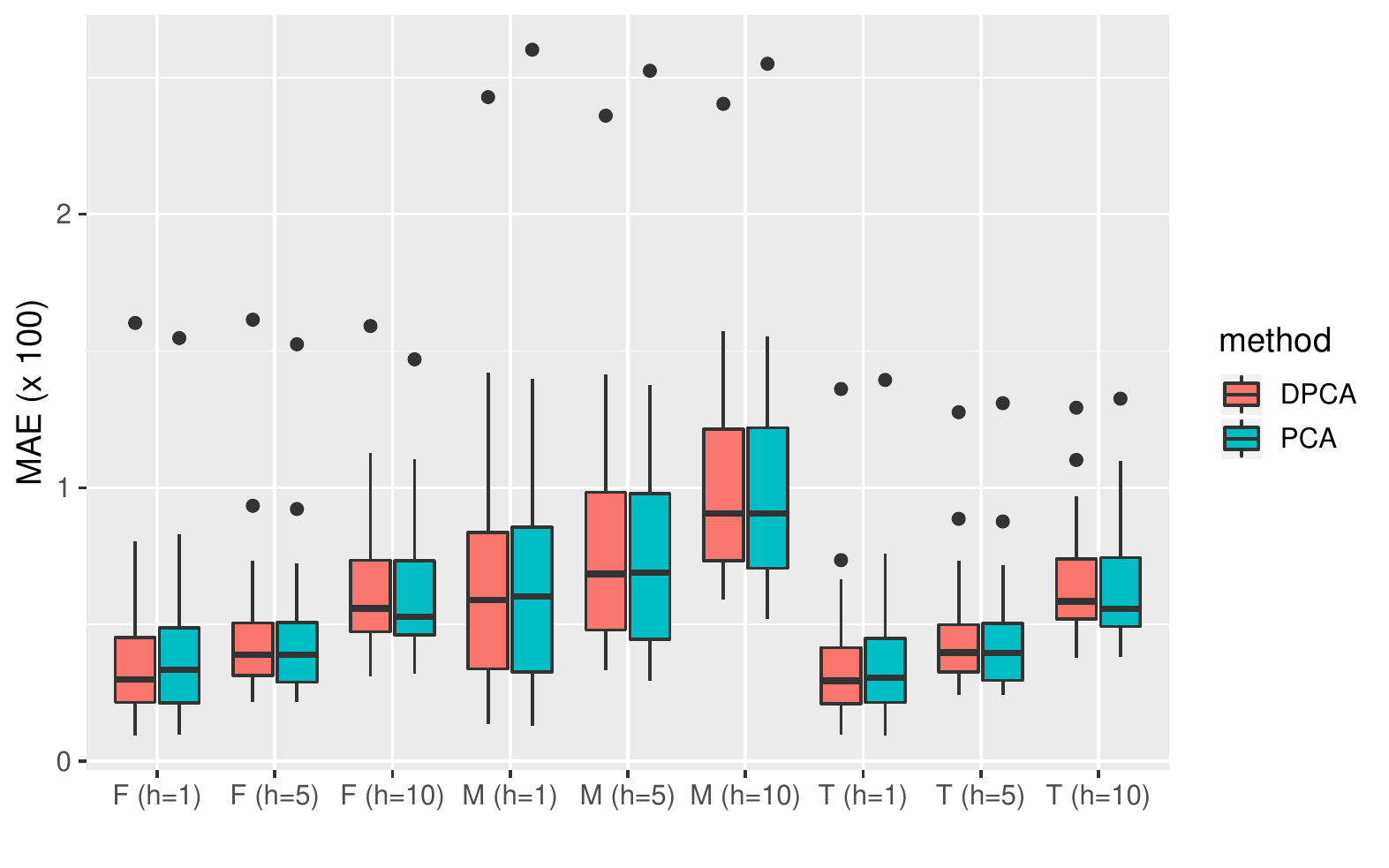}
\\
\includegraphics[width=\textwidth]{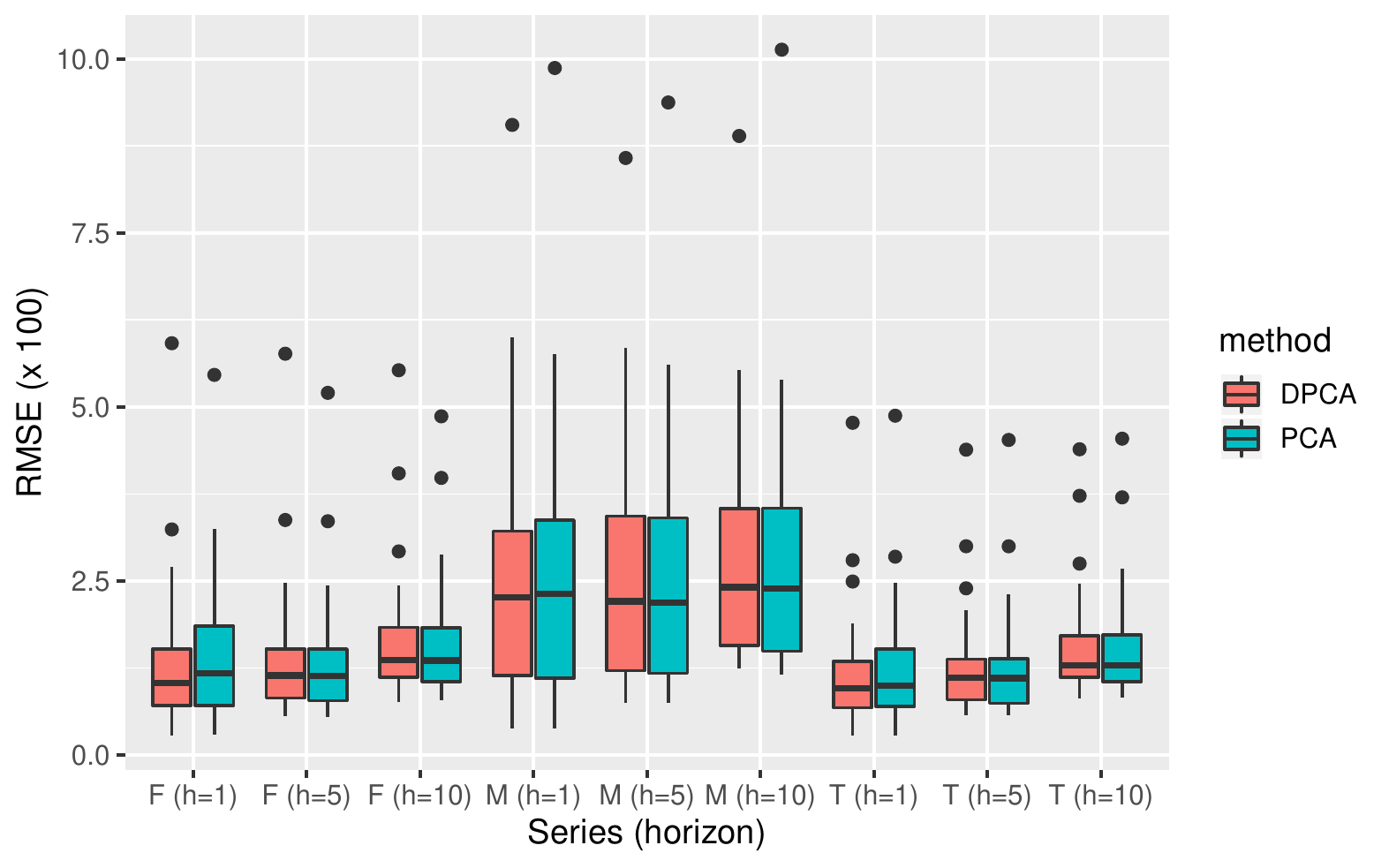}
\caption{Comparison of one-step-ahead, five-step-ahead and 10-step-ahead point forecast errors between the dynamic and static principal component regression using the LC method without centering (i.e., \citeauthor{HR12}'s \citeyearpar{HR12} method). $h$ denotes forecast horizon.}\label{fig:HR12_tab}
\end{figure}

In \cite{HR12}, the Lee-Carter (LC) method does not include the mean term, since their Generalized Linear Model approach uses an iterative fitting algorithm to estimate parameters by minimizing a deviance criterion. In Figure~\ref{fig:HR12_tab}, we present the point forecast results for the one-step-ahead, five-step-ahead, and 10-step-ahead forecasts. The results are similar to the LC method with the mean term given in Table~\ref{tab:point_accuracy} of the manuscript and results given in \mbox{Appendix C}. 

\section*{Appendix C}

\begin{figure}[!htbp]
\centering
\includegraphics[width=8.7cm]{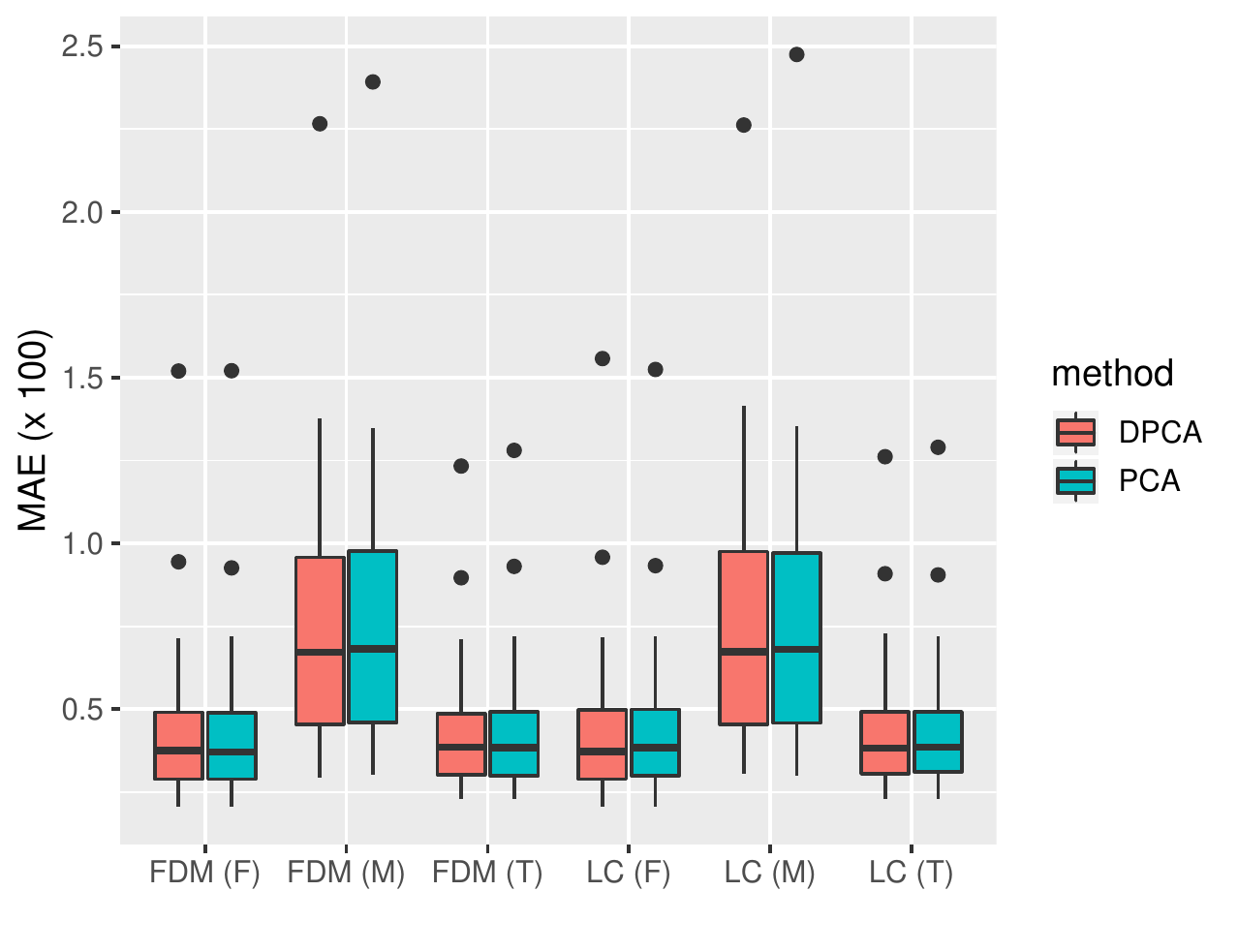}
\quad
\includegraphics[width=8.7cm]{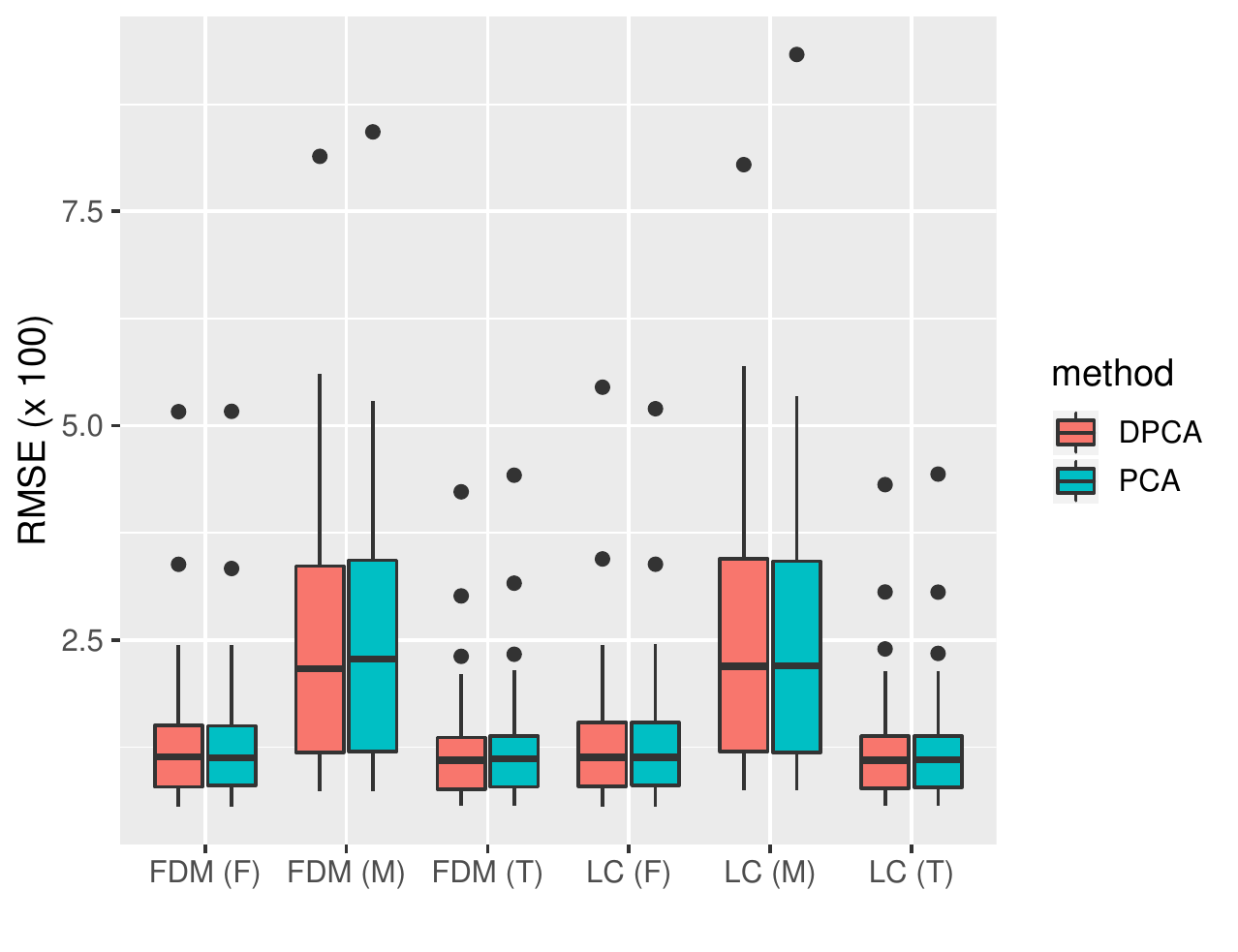}
\\
\includegraphics[width=8.7cm]{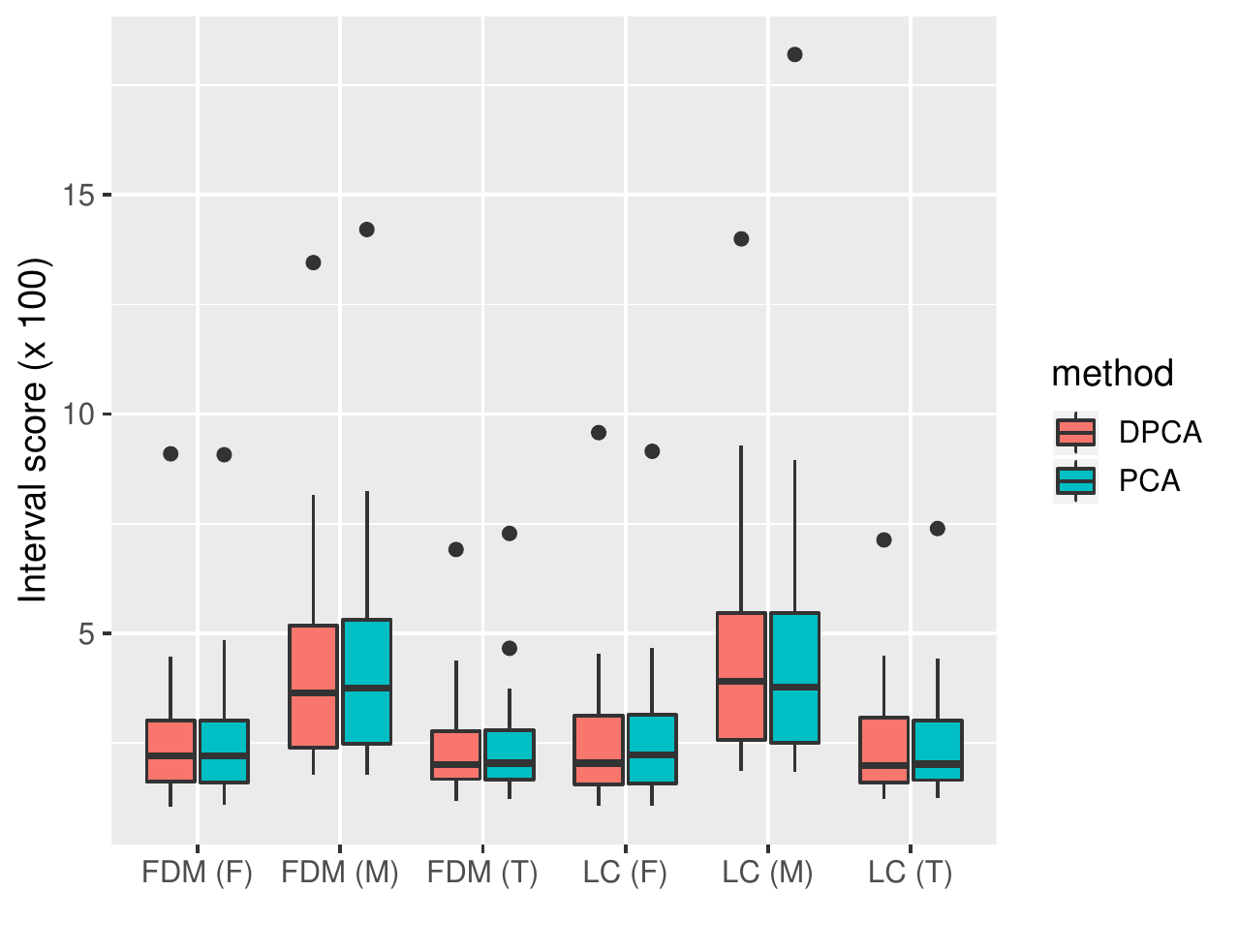}
\quad
\includegraphics[width=8.7cm]{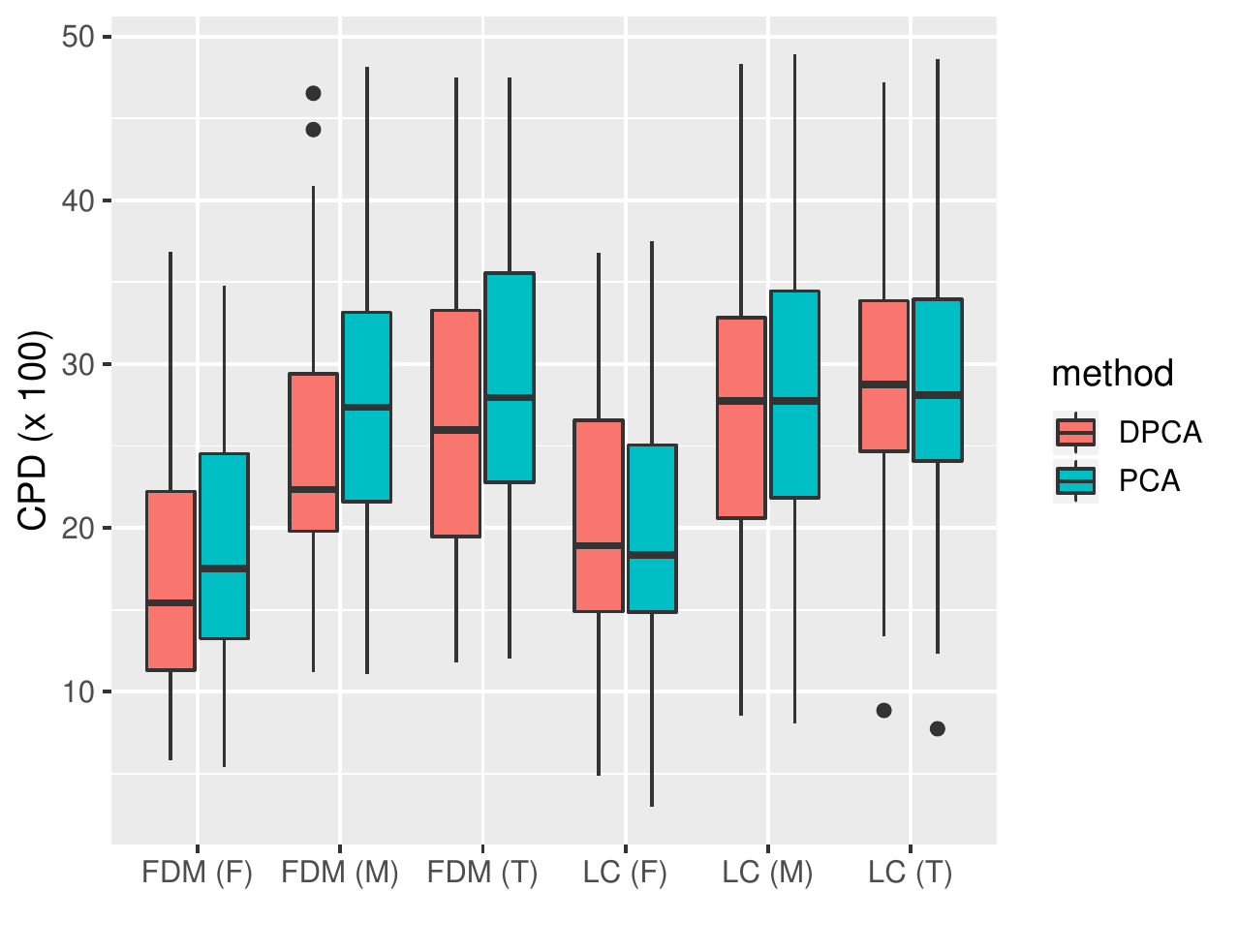}
\caption{Comparison of five-step-ahead point and interval forecast errors between the dynamic and static principal component regression using the functional time-series method and LC method with centering.}\label{fig:boxplot_h_5}
\end{figure}

\begin{figure}[!htbp]
\centering
\includegraphics[width=8.7cm]{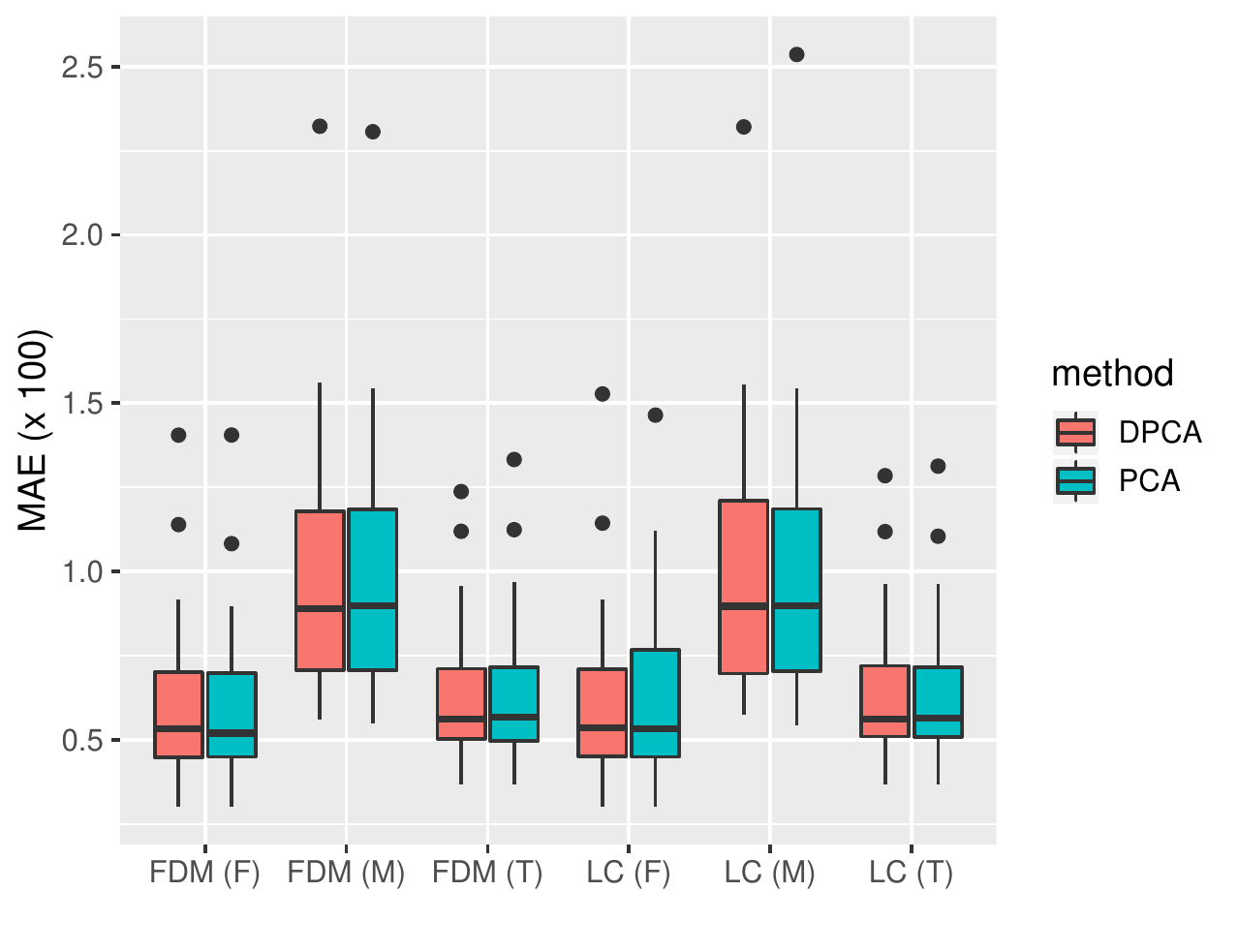}
\quad
\includegraphics[width=8.7cm]{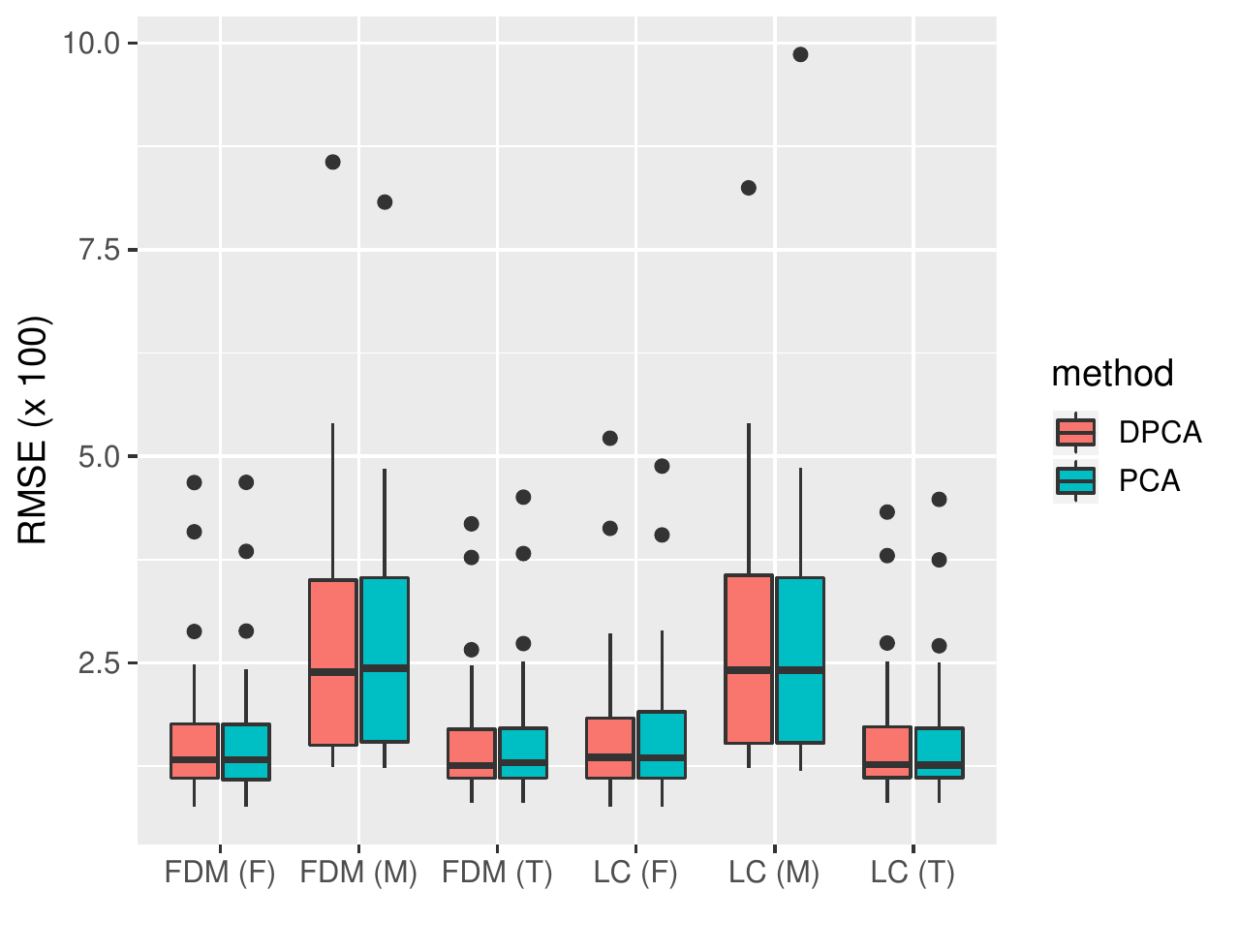}
\\
\includegraphics[width=8.7cm]{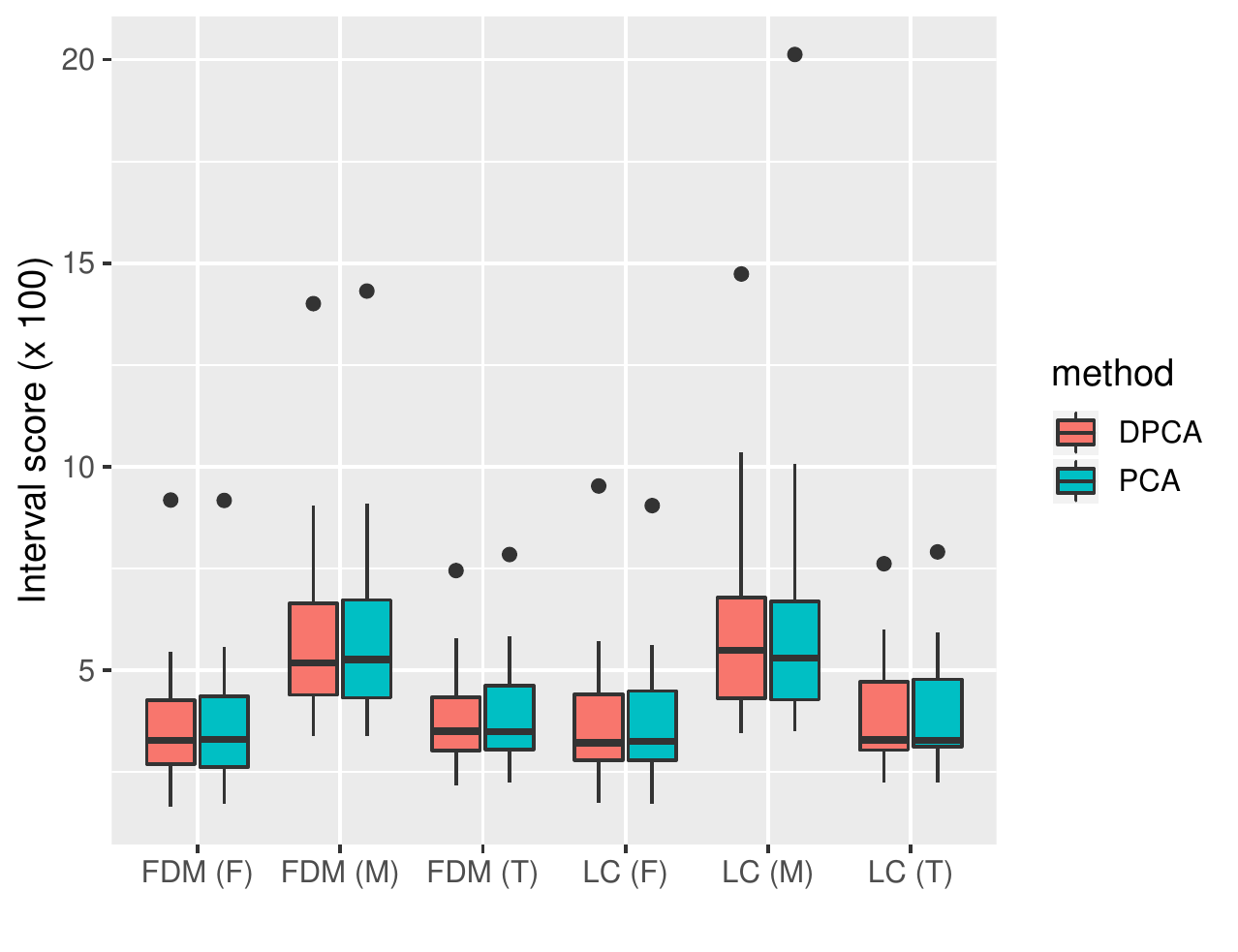}
\quad
\includegraphics[width=8.7cm]{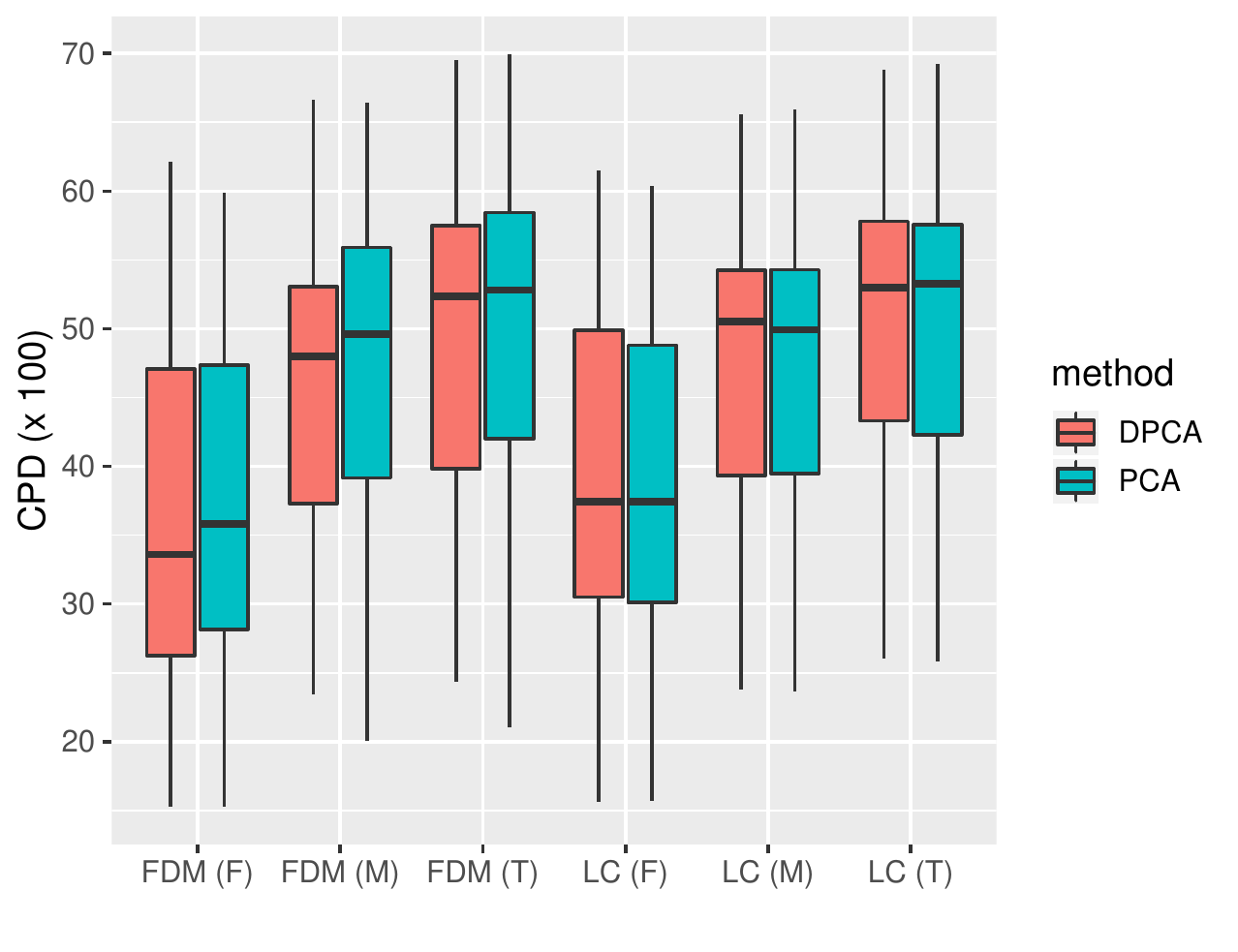}
\caption{Comparison of 10-step-ahead point and interval forecast errors between the dynamic and static principal component regression using the functional time-series method and LC method with centering.}\label{fig:boxplot_h_10}
\end{figure}

Using the LC and functional time-series methods, these longer-horizon forecast results are reported in Figure~\ref{fig:boxplot_h_5} for $h=5$. For $h=10$, these results are reported in Figure~\ref{fig:boxplot_h_10}. As measured by MAE, RMSE and interval score, there are marginal difference between the DPCA and static PCA, although the maximum forecast error of the DPCA is often smaller than that of the static PCA. As measured by CPD, DPCA often outperforms the static PCA.

\newpage
\bibliographystyle{agsm}
\bibliography{DFPCA}

\end{document}